%% file: sgrcdisk.tex
\documentclass[trackchanges,twocolumn]{aastex701}

\usepackage{CJKutf8}
\usepackage{amsmath}

\newcommand{\msol}{\mbox{$M_\odot$}}

\begin{document}
\begin{CJK}{UTF8}{gbsn}

\title{The Keplerian disk, envelope, and streamers surrounding an early O-type protostar in the Sagittarius C cloud of the Central Molecular Zone}

\input{authors}

\begin{abstract}
Disk-mediated accretion is central to theories of massive star formation, setting the initial conditions for their evolution. Yet observations of Keplerian disks around early O-type protostars remain scarce, as they are often blended into complex surrounding structures. We report ALMA Band 6 observations (300~au resolution) of an accretion disk surrounding a high-mass protostar in the Sagittarius C (Sgr~C) cloud in the Central Molecular Zone (CMZ) around the Galactic Center. We identify spectral lines and analyze the spatial distribution of the emission of the complex organic molecules. We use a dynamical model with an inner Keplerian disk and an outer free-fall envelope to fit the three-dimensional position-position-velocity data of the stacked CH$_3$OCHO molecular lines and constrain the mass of the central protostar to be $\sim40^{+2}_{-3}~\msol{}$. The fitting results additionally show that the disk has a centrifugal radius at about 1300~au. Considering the infall velocity, radius, and mass of the envelope, we estimate the accretion rate from the envelope onto the disk to be $\sim7\times 10^{-3}\ M_{\odot}\,\mathrm{yr^{-1}}$. We also identify spiral-like structures in the disk that can be described by free-falling streamers. Our results highlight the critical role of accretion disks and streamers in the mass accumulation of early O-type stars in the CMZ.
\end{abstract}


\section{Introduction}
\label{sec:intro}

The initial mass of a star greatly impacts its evolutionary path. Understanding how stars, particularly the high-mass ones ($M_{\star} > 8\ M_{\odot}$), acquire their masses is a central challenge in modern astrophysics \citep{McKee2007,Motte2018,Beuther2025}. A long-standing theoretical hurdle is the intense radiation pressure from a luminous O/B-type protostar, which would halt the inflow of accreting material and limit the final stellar mass \citep{kahn1974, Wolfire1987}. Disk-mediated accretion, which channels material non-spherically onto the protostar, is the most widely accepted solution to this ``radiation pressure problem'' \citep{Bonnell2006,Keto2007,Krumholz2009}. The detection and characterization of accretion disks around high-mass protostars are therefore crucial for understanding how they assemble their mass and acquire their final angular momentum.

With the revolutionary sensitivity and angular resolution of the Atacama Large Millimeter/submillimeter Array (ALMA), observations have established the presence of disks around high-mass protostars. The current observational landscape reveals at least two distinct regimes (\citealp{Beltran2020} and references therein). For early B-type or late O-type stars ($M_{\star} \lesssim 20\ M_{\odot}$), observations have resolved numerous rotationally supported {(quasi\nobreakdash-)Keplerian} disks on scales of hundreds to thousands of au \citep[e.g.,][]{Maud2017,Ginsburg2018,Sanna2019,Zapata2019}. These structures appear to be scaled-up versions of disks surrounding low-mass protostars. For the most massive and luminous early O-type protostars ($M_{\star} \gtrsim 30\ M_{\odot}$), Keplerian disks remain rare \citep{Beltran2020}. Recent studies of this type have yielded a handful of candidate Keplerian disks around O-type (proto)stars \citep[e.g.,][]{Johnston2015,Ilee2018,Maud2019}. The study by \citet{Olguin2026} offers a uniform analysis of a well-characterized high-mass protostellar disk sample, featuring two prominent cases of 45 $M_{\odot}$ stars. However, observations of these candidate Keplerian disks at high resolution often reveal large ($\sim 10^3$--$10^4$ au), massive (up to $\sim 100\ M_{\odot}$), and in some cases asymmetric rotating structures or ``toroids'' \citep[e.g.,][]{Sollins2005,Cesaroni2017,Liu2017}. The kinematics of these structures are typically non-Keplerian, suggesting that they may be transient, gravitationally unstable structures, or pseudo-disks dominated by large-scale infall, rather than stable disks supported by rotation. Furthermore, several high-resolution studies suggesting that in some high-mass systems, true disks may be exceptionally compact or confused with surrounding structures \citep[e.g.,][]{Goddi2020, Olguin2025}. Consequently, characterizing these systems now requires using high-resolution kinematics to disentangle a complex hierarchy of components, including large-scale envelopes, infall streamers, pseudo-disks, and potentially small Keplerian disks embedded deep within \citep{Zhang2019,Johnston2020a,Beltran2022,Xiaofeng2025,Olguin2025}.

Here, we present the case of a massive disk/envelope system surrounding a high-mass protostar G359.44$-$0.102 ($\mathrm{R.A.} = 17^{\rm h}44^{\rm m}40.153^{\rm s}, \mathrm{Decl.}=-29\arcdeg 28\arcmin 12.876\arcsec$), which locates in the Sagittarius C molecular cloud in the Central Molecular Zone (CMZ) at a distance of 8.3~kpc \citep{GRAVITY2019}, hereafter referred to as the Sgr C disk. The CMZ is a massive reservoir of molecular gas surrounding the Galactic Center \citep{Morris1996,Henshaw2023}. Despite its high gas densities, its global star formation efficiency is suppressed by a factor of $\sim$10 compared to predictions from empirical dense gas-star formation correlations \citep{Kauffmann2017,Barnes2017}, with only a few regions of known active massive star formation including the Sagittarius C cloud \citep{LuXing2019,Battersby2025}. 
The Sgr~C disk surrounds the high-mass protostar G359.44$-$0.102 that has a mass of $\gtrsim 30\ M_{\odot}$ based on the assumption of Keplerian rotation in the disk, and drives powerful outflows \citep{Luxing2021,LuXing2022}. As a candidate for a Keplerian disk around an early O-type protostar, the Sgr~C disk provides a chance to test models of mass accretion in extreme environments.

The remainder of this paper is organized as follows. \autoref{sec:obs} describes the ALMA Band 6 observations and data reduction. In \autoref{sec:result}, we identify molecular lines, derive the physical properties of the disk, and employ Non-negative Matrix Factorization (NMF) to analyze chemical segregation. We then apply a composite dynamic model that combines a Keplerian disk and an Ulrich envelope to fit the position-position-velocity (PPV) data. \autoref{sec:discussion} discusses the observed spiral structures, the high accretion rate, and the evolutionary state of the high-mass protostar. Our main findings are summarized in \autoref{sec:summary}.

\section{Observations}\label{sec:obs}

The ALMA 12-m array observations towards the Sgr~C region were carried out in a Cycle 6 program (ID: 2018.1.00641.S; PI: Xing Lu) in five epochs on 2019 June 5 and July 14--15 with the nominal configuration of C43-8, as part of the Broadband ALMA Longbaseline Look-in at Accretion Disks in the CMZ (BALLAD) project. For Sgr~C, the pointing center was $\alpha$(J200) = $17^{\rm h}44^{\rm m}40.37^{\rm s}$, $\delta$(J2000) = $-29\arcdeg 28\arcmin 14.79\arcsec$. This position was offset from the continuum emission peak of the Sgr~C disk to simultaneously cover another target in the region.

The correlators were tuned to Band 6 ($1.3$~mm) with four spectral windows (SPWs), which we denote as SPW0 to SPW4. SPW0 covers the frequency range 216.95--218.83~GHz (centered at 217.105~GHz); SPW1 covers 220.02--220.95~GHz (centered at 220.747~GHz); SPW2 covers 231.15--233.02~GHz (centered at 231.901~GHz); and SPW3 covers 233.15--235.02~GHz (centered at 234.252~GHz). These SPWs cover several molecular lines, including complex organic molecules (COMs) such as CH$_3$OCHO, as well as CH$_3$CN and its isotopologues. Between 42 and 48 operational 12-m antennas were in the array during the observations, with baseline lengths ranging from 83~m to 15,238~m. The maximum recoverable scale (MRS) is 0\farcs{7}, as determined by the fifth percentile shortest baselines in the array, and the full-width at half-maximum (FWHM) primary beam size (field of view) is $26\arcsec$ ($\sim$1.05~pc). The total on-source integration time for the Sgr~C region was about 64.6~minutes. The quasar J1924$-$2914 was used for absolute flux and bandpass calibration, and J1744$-$3116 was used for phase calibration.

We calibrated the raw visibilities using the standard pipeline in CASA version 5.4.0. For self-calibration and imaging, we used CASA version 5.6.1. Line-free channels were determined by examining the spectra in the visibility domain. Using these channels, we reconstructed the continuum, and ran self-calibration to improve the dynamic range for the five epochs altogether. We performed six rounds of phase-only self calibration, with solution intervals progressively decreasing from 1 scan (about 54.4~second) down to 9.072 seconds. This was followed by one round of amplitude self calibration using a solution interval of 1 scan.
After self calibration, the dynamic range was doubled to roughly 700. The final continuum image achieved a $1\sigma$ rms noise of $18~\mu\text{Jy beam}^{-1}$ with a peak intensity of $12.5~\text{mJy beam}^{-1}$.

Then we applied the self-calibration solutions to both the continuum data and the spectral line data. The continuum data from the five epochs were combined and imaged using the TCLEAN task, with a Briggs weighting parameter of 0.5, a gridder mode of `mosaic', and an image size of 6750 pixels with a pixel size of 0\farcs{007} that leads to an image angular size of 47\farcs{25}. The multiscale multifrequency synthesis algorithm was employed, with multiscale parameters of [0, 5, 15, 50] times the pixel size. The resulting synthesized beam size for the continuum is $0.043\arcsec \times 0.030\arcsec$ with a position angle of $72.75\arcdeg$. This corresponds to a linear scale of $ 350\ \text{au} \times 240\ \text{au}$, assuming a distance of 8.3~kpc to the Galactic Center \citep{GRAVITY2019}. The spectral lines were imaged using the same parameters. The synthesized beam size for the spectral lines is on average $0.053\arcsec \times 0.032\arcsec$ with a position angle of $66.15\arcdeg$ but slightly varies among SPWs. The resulting spectral line cubes achieved $1\sigma$ rms noise of $0.8~\text{mJy beam}^{-1}$ (equivalent to $\sim$11~K) per $0.67~\text{km s}^{-1}$ channel width. More details on the observational setup and data reduction are described in \citet{LuXing2022} and \citet{Zhang2025}.

\section{Results}\label{sec:result}
\begin{figure*}[!thpb]
\centering
\includegraphics[width=1\textwidth]{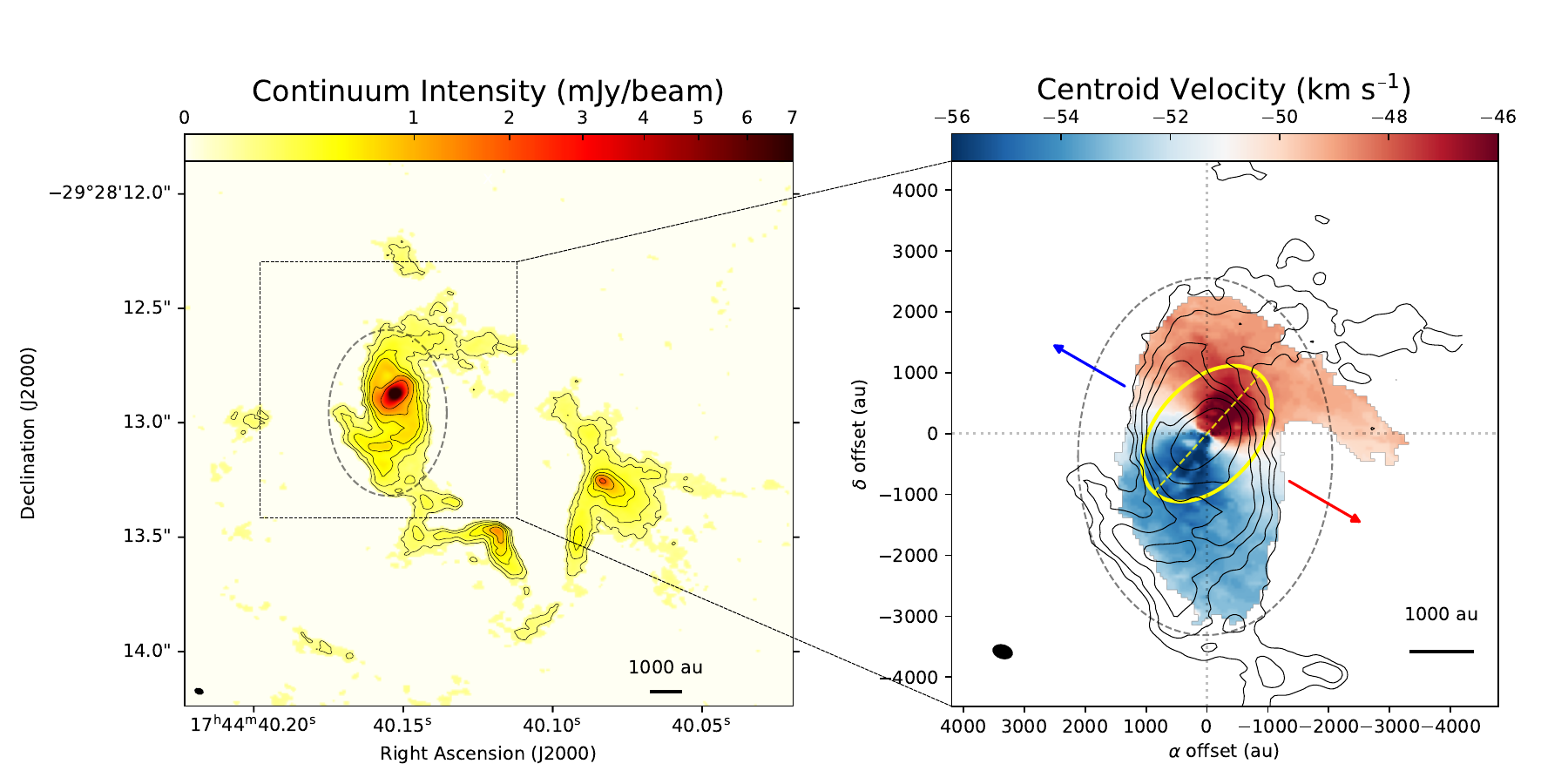}
\caption{
\textbf{Left:} ALMA 1.3\,mm continuum intensity map of the Sgr C disk. The black contours indicate intensity levels at [5, 10, 20, 30, 50, and 100]$\times\sigma$, where the RMS noise $\sigma$ is $18~\mu\mathrm{Jy\,beam^{-1}}$. The dashed ellipse denotes the region over which the spectra in \autoref{fig:line identification} were averaged. The dashed rectangle indicates the field of view shown in the right panel. 
\textbf{Right:} intensity-weighted velocity map of the stacked CH$_3$OCHO emission. 
The color scale shows the LSR velocity, revealing a velocity gradient consistent with rotation. Black contours depict the continuum emission as in the left panel. The blue and red arrows show the directions of the outflow seen in SiO emission in lower-resolution data \citep{Luxing2021}. The yellow ellipse outlines the best-fit projected Keplerian disk geometry, characterized by a radius of $\sim1300$~au, a position angle of $42^{\circ}$ (measured from North to West), and an inclination angle of $50^{\circ}$, where $0^{\circ}$ means face-on and $90^{\circ}$ edge-on. The yellow dashed line is the major axis of the ellipse.
}
\label{fig:cont and moment1}
\end{figure*}

\subsection{Line identification}\label{subsec: line identification}

\begin{figure*}[!thpb]
\centering
\includegraphics[width=1.3\textwidth, angle=-90]{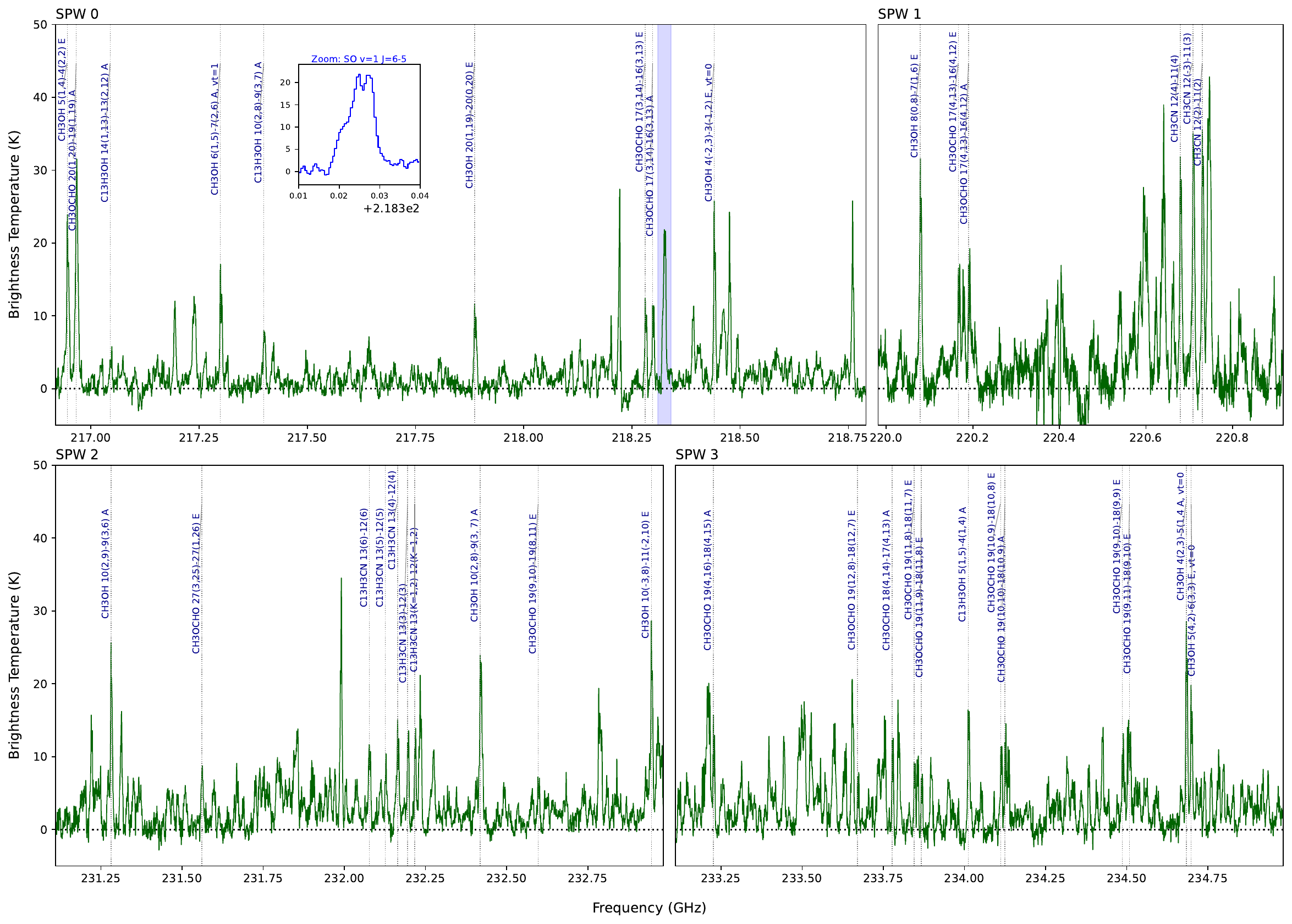} \\
\caption{Spectral line identification results extracted from the four spectral windows averaged over the inner disk region denoted by the dashed ellipse in \autoref{fig:cont and moment1}. Only the non-blended lines used in the NMF analysis or velocity stacking are labeled. The inset provides a zoomed-in view of the SO v=1~$\rm J=6$--5 emission.}
\label{fig:line identification}
\end{figure*}

We integrated over the area of the black dashed ellipse in \autoref{fig:cont and moment1} for the four observed SPWs to obtain the mean spectra of the Sgr~C disk in these frequency ranges. The resulting spectra, presented in \autoref{fig:line identification}, reveal a rich forest of spectral lines. Lines were identified using Spectuner \citep{Qiu2025}, a tool for automated spectral line analysis that implements a one-dimensional local thermodynamic equilibrium (LTE) spectral model. The results demonstrate that the disk is abundant in COMs, such as CH$_3$OH, $^{13}$CH$_3$OH, CH$_3$OCHO, CH$_3$OCH$_3$, CH$_3$CN, $^{13}$CH$_3$CN, and C$_2$H$_5$CN. In addition, emission from other species is detected, including H$_2$CO, SO, and CS. In addition to the transitions labeled in \autoref{fig:line identification}, numerous other transitions originating from the aforementioned species are present. However, they are not labeled due to significant line blending and there are excluded from the subsequent analysis.

The inset in \autoref{fig:line identification} displays the SO v=1 $\rm J=6$--5 line, revealing a double-peaked profile, which is characteristic of the vast majority of identified molecular lines. Such double-peaked profiles are consistent with the expected spatially integrated emission from a rotating disk \citep{Smak1982}, suggesting a possible origin from the rotation of the Sgr~C disk. In \autoref{tab:lines}, we listed the 52 identified non-blended spectral lines, along with their parameters, including their quantum numbers, upper state energies ($E_u$), rest frequencies, Einstein A coefficients ($\log_{10}(A_{ij})$), and whether they are used for the NMF analysis (see \autoref{subsec:NMF}). 

\begin{deluxetable*}{llccccc}
\tabletypesize{\scriptsize}
\tablecaption{Identified Spectral Line Parameters \label{tab:lines}}
\tablewidth{0pt}
\tablehead{
\colhead{Molecule} & \colhead{Vibrational state} & \colhead{Transition} & \colhead{$E_u$} & \colhead{Rest Freq.} & \colhead{$\log_{10}(A_{ij})$} & \colhead{NMF}  \\
\colhead{} & \colhead{} & \colhead{} & \colhead{(K)} & \colhead{(MHz)} & \colhead{(s$^{-1}$)} & \colhead{}
}
\startdata
$^{13}$CH$_3$CN &  & 13(6)--12(6) & 335.53 & 232077.197 & $-$3.074 & Y \\
$^{13}$CH$_3$CN &  & 13(5)--12(5) & 256.88 & 232126.006 & $-$3.039 & Y  \\
$^{13}$CH$_3$CN &  & 13(4)--12(4) & 191.89 & 232164.355 & $-$3.012 & Y  \\
$^{13}$CH$_3$CN &  & 13(3)--12(3) & 142.08 & 232194.888 & $-$2.993 & Y  \\
$^{13}$CH$_3$CN &  & 13(2)--12(2) & 106.5 & 232216.706 & $-$2.979 & Y  \\
$^{13}$CH$_3$CN &  & 13(1)--12(1) & 85.14 & 232229.801 & $-$2.971 & N  \\
$^{13}$CH$_3$OH & vt=0 & 14(1,13)--13(2,12) A & 254.25 & 217044.616 & $-$4.624 & Y \\
$^{13}$CH$_3$OH & vt=0 & 10(2,8)--9(3,7) A & 162.41 & 217399.55 & $-$4.816 & Y \\
$^{13}$CH$_3$OH & vt=0 & 5(1,5)--4(1,4) A & 48.25 & 234011.58 & $-$4.278 & Y \\
$^{13}$CS & v=0 & 5--4 & 33.29 & 231220.672 & $-$3.601 & N \\
C$_2$H$_5$CN & v=0 & 26(1,25)--25(1,24) & 153.42 & 231310.42 & $-$2.982 & Y \\
C$_2$H$_5$CN & v=0 & 26(5,22)--25(5,21) & 178.85 & 233443.1 & $-$2.985 & Y \\
C$_2$H$_5$CN & v=0 & 22(4,18)--23(0,23) & 126.73 & 231987.114 & $-$8.003 & Y \\
CH$_3$CN & v=0 & 12(4)--11(4) & 183.15 & 220679.228 & $-$3.089 & Y\\
CH$_3$CN & v=0 & 12($-$3)--11(3) F=11--10 & 133.16 & 220709.017 & $-$3.062 & N \\
CH$_3$CN & v=0 & 12(2)--11(2) & 97.44 & 220730.261 & $-$3.208 & N \\
CH$_3$OCH$_3$ & v=0 & 22(4,19)--22(3,20) AA & 253.41 & 217193.153 & $-$3.265 & Y \\
CH$_3$OCHO & v=0 & 27(3,25)--27(1,26) E & 221.73 & 231561.307 & $-$5.394 & Y \\
CH$_3$OCHO & v=0 & 20(1,20)--19(1,19) A & 111.48 & 216965.938 & $-$3.822 & Y \\
CH$_3$OCHO & v=0 & 19(4,16)--18(4,15) A & 123.25 & 233226.782 & $-$3.742 & Y \\
CH$_3$OCHO & v=0 & 19(12,8)--18(12,7) E & 207.6 & 233670.944 & $-$3.934 & Y \\
CH$_3$OCHO & v=0 & 19(11,8)--18(11,7) E & 192.39 & 233845.126 & $-$3.89 & Y \\
CH$_3$OCHO & v=0 & 19(11,9)--18(11,8) A & 192.39 & 233854.263 & $-$3.89 & Y \\
CH$_3$OCHO & v=0 & 19(11,9)--18(11,8) E & 192.38 & 233867.128 & $-$3.89 & Y \\
CH$_3$OCHO & v=0 & 19(10,9)--18(10,8) E & 178.51 & 234112.25 & $-$3.853 & Y \\
CH$_3$OCHO & v=0 & 19(10,10)--18(10,9) A & 178.5 & 234124.88 & $-$3.853 & Y \\
CH$_3$OCHO & v=0 & 19(10,10)--18(10,9) E & 178.49 & 234134.59 & $-$3.853 & Y \\
CH$_3$OCHO & v=0 & 19(9,10)--18(9,9) E & 165.98 & 234486.41 & $-$3.821 & N \\
CH$_3$OCHO & v=0 & 19(9,10)--19(8,11) E & 165.98 & 232597.272 & $-$4.817 & Y \\
CH$_3$OCHO & v=0 & 19(9,11)--18(9,10) A & 165.97 & 234502.25 & $-$3.82  & N\\
CH$_3$OCHO & v=0 & 19(9,11)--18(9,10) E & 165.97 & 234508.55 & $-$3.82 & N \\
CH$_3$OCHO & v=0 & 18(4,14)--17(4,13) A & 114.36 & 233777.515 & $-$3.735 & Y \\
CH$_3$OCHO & v=0 & 17(3,14)--16(3,13) E & 99.73 & 218280.83 & $-$3.821 & Y \\
CH$_3$OCHO & v=0 & 17(3,14)--16(3,13) A & 99.72 & 218297.866 & $-$3.821 & Y \\
CH$_3$OCHO & v=0 & 17(4,13)--16(4,12) E & 103.15 & 220166.888 & $-$3.817 & Y \\
CH$_3$OCHO & v=0 & 17(4,13)--16(4,12) A & 103.14 & 220190.268 & $-$3.816 & Y \\
CH$_3$OH & vt=0--2 & 5(1,4)--4(2,2) E & 55.87 & 216945.559 & $-$4.916 & Y \\
CH$_3$OH & vt=0--2 & 20(1,19)--20(0,20) E & 508.38 & 217886.39 & $-$4.471 & Y \\
CH$_3$OH & vt=0--2 & 4($-$2,3)--3($-$1,2) E, vt=0 & 45.46 & 218440.063 & $-$4.329 & N \\
CH$_3$OH & vt=0--2 & 8(0,8)--7(1,6) E & 96.61 & 220078.49 & $-$4.599 & Y \\
CH$_3$OH & vt=0--2 & 10(2,9)--9(3,6) A & 165.35 & 231281.15 & $-$4.737 & Y  \\
CH$_3$OH & vt=0--2 & 10(2,8)--9(3,7) A & 165.4 & 232418.59 & $-$4.728 & Y  \\
CH$_3$OH & vt=0--2 & 10($-$3,8)--11($-$2,10) E & 190.37 & 232945.835 & $-$4.672 & Y \\
CH$_3$OH & vt=0--2 & 18(3,15)--17(4,14) A & 446.58 & 233795.799 & $-$4.657 & Y  \\
CH$_3$OH & vt=0--2 & 4(2,3)--5(1,4) A, vt=0 & 60.92 & 234683.451 & $-$4.727 & Y \\
CH$_3$OH & vt=0--2 & 5(4,2)--6(3,3) E, vt=0 & 122.72 & 234698.519 & $-$5.198 & Y  \\
CH$_3$OH & v12=1 & 6(1,5)--7(2,6) A, vt=1 & 373.93 & 217299.202 & $-$4.368 & Y  \\
DCN & v=0 & J=3--2, F=2--2 & 20.85 & 217238.631 & $-$4.243 & Y  \\
H$_2$CO &  & 3(0,3)--2(0,2) & 20.96 & 218222.192 & $-$3.55 & N  \\
H$_2$CO &  & 3(2,2)--2(2,1) & 68.09 & 218475.642 & $-$3.595 & N  \\
H$_2$CO &  & 3(2,1)--2(2,0) & 68.11 & 218760.066 & $-$3.802 & N  \\
SO & v=1 & 6(5)--5(4) & 1633.95 & 218323.858 & $-$3.876 & N  \\
\enddata
\tablecomments{This table lists the parameters of the spectral lines identified in the data. Column 1: Molecule. Column 2: Vibrational or torsional energy state. Column 3: Quantum numbers of the transition. Column 4: Upper state energy. Column 5: Rest frequency. Column 6: Einstein A coefficient. Column 7: Whether being used in the NMF analysis. Molecular line information is taken from the CDMS database \citep{CDMS2001}.
}
\end{deluxetable*}

\subsection{Rotational temperature and disk mass}\label{subsec: temperature and mass}

\begin{figure*}[!thpb]
\centering
\includegraphics*[width=0.45\textwidth]{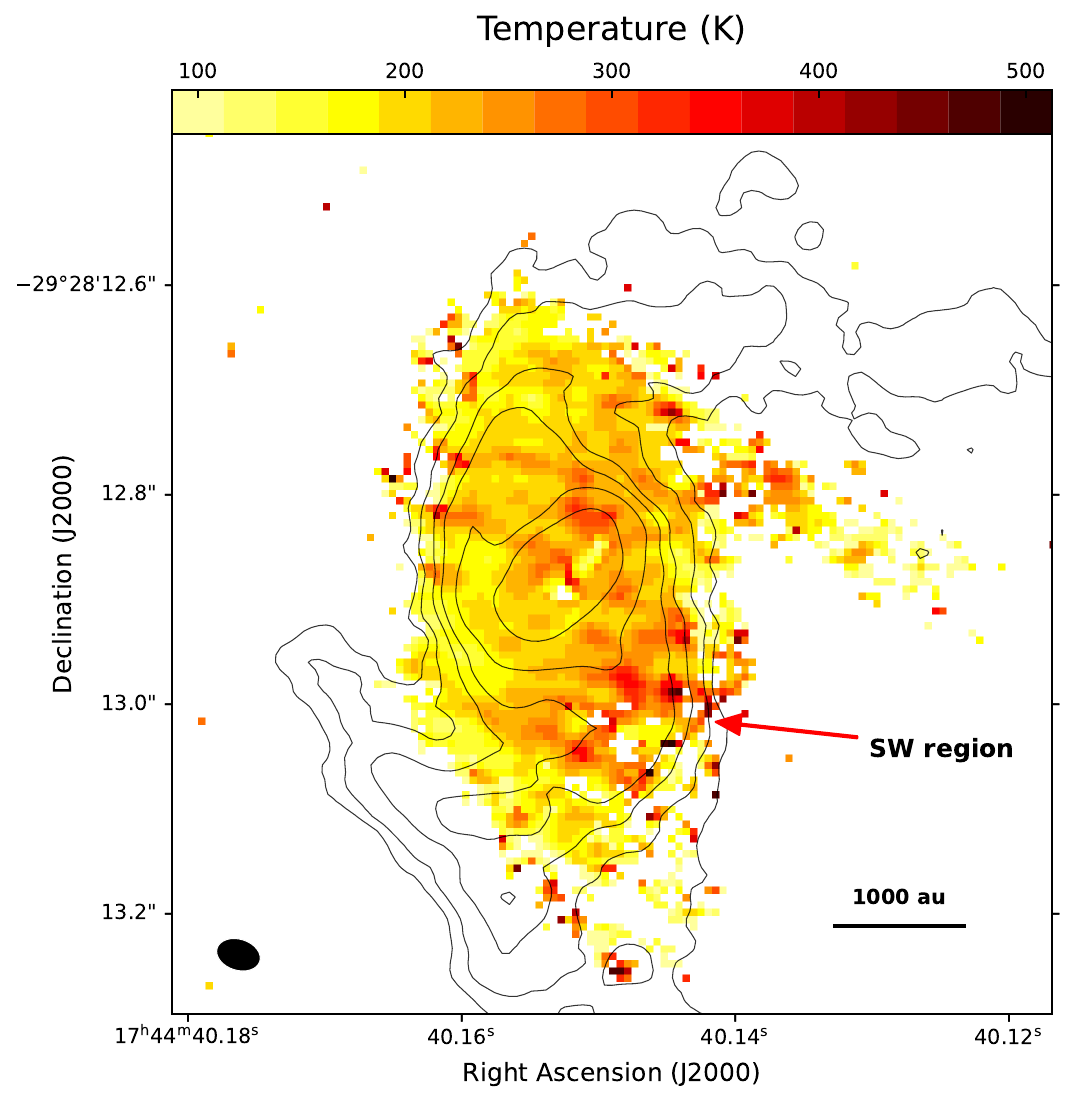}
\includegraphics*[width=0.45\textwidth]{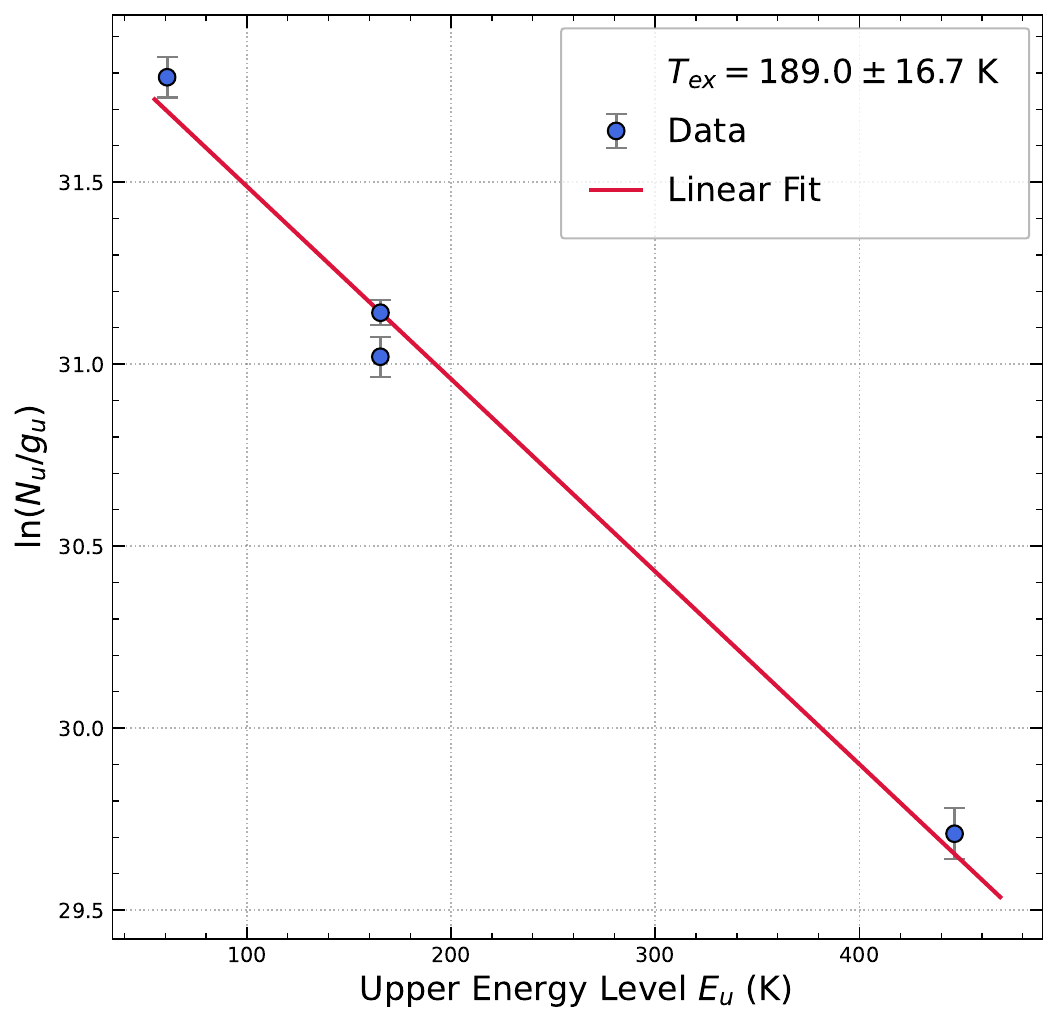}
\caption{\textbf{Left}: Spatial distribution of the CH$_3$OH rotational temperature within the Sgr~C disk. The black contours indicate 1.3 mm continuum emission levels identical to \autoref{fig:cont and moment1}. \textbf{Right}: Rotational diagram of CH$_3$OH transitions, using integrated intensities detected above 3$\sigma$. The solid line represents the linear fit, yielding a temperature of 189 K.}
\label{fig:rotational temperature}
\end{figure*}

\autoref{fig:rotational temperature} presents the spatial distribution of the rotational temperature derived for the Sgr~C disk. We determined the excitation temperature ($T_{\rm ex}$) and total column density ($N_{\rm tot}$) using the rotational diagram \citep{Goldsmith1999}. This approach assumes local thermodynamic equilibrium (LTE) and that the emission is optically thin and fills the beam. Under these conditions, the linearized Boltzmann equation is given by:
\begin{equation}
\ln\left(\frac{N_u}{g_u}\right) = \ln\left(\frac{N_{\rm tot}}{Q(T_{\rm ex})}\right) - \frac{E_u}{T_{\rm ex}},
\end{equation}
where $g_u$ is the degeneracy of the upper state, $Q(T_{\rm ex})$ is the partition function, and $E_u$ is the upper-level energy in units of Kelvin. The upper-level column density, $N_u$, is related to the velocity-integrated intensity ($W = \int T_A dv$, in $\mathrm{K\ km\ s^{-1}}$) via:
\begin{equation}
N_u = \frac{8\pi k_B \nu^2}{h c^3 A_{ul}} W,
\end{equation}
where $\nu$ is the rest frequency, $A_{ul}$ is the Einstein $A$-coefficient, $k_B$ is the Boltzmann constant, and $h$ is the Planck constant. A linear regression of $\ln(N_u/g_u)$ versus $E_u$ yields $T_{\rm ex}$ from the slope ($-1/T_{\rm ex}$) and $N_{\rm tot}$ from the intercept.

The temperature map of the Sgr C disk was constructed using four $\mathrm{CH_3OH}$ transitions: $10(2,9)$--$9(3,6)$\,A, $10(2,8)$--$9(3,7)$\,A, $18(3,15)$--$17(4,14)$\,A, and $4(2,3)$--$5(1,4)$\,A. The derived temperatures range from 100 to 260 K, with an interquartile range of [133, 172]~K, which are consistent with the temperatures derived using XCLASS to fit $^{13}$CH$_3$CN \citep{LuXing2022}. Spurious high-temperature pixels are observed at the disk boundaries; these likely arise from the low signal-to-noise ratio (SNR) in the outer regions, which compromises the reliability of the temperature fit. We noted that in the southwestern (SW) region, the derived temperatures are also anomalously high, and in certain pixels, the fit fails to converge. The observed high-temperature structure appears to extend along the Northeast-Southwest (NE-SW) axis. This spatial distribution is consistent with the outflow associated with the Sgr C disk. Since CH$_3$OH is highly sensitive to shocked gas, elevated temperatures in these regions likely reflect the impact of outflows on the surrounding envelope. Meanwhile, the emission of the four $\mathrm{CH_3OH}$ transitions shows a low SNR in the SW region, which may result in large uncertainties in the derived temperatures (see \autoref{app:moment0}).

In addition to the temperature map, we derived an average rotational temperature of 189~K for the whole disk, by fitting the rotational diagram using the integrated line fluxes above 3$\sigma$ (\autoref{fig:rotational temperature} right panel). 

We estimated the total gas mass of the disk based on the 1.3 mm thermal dust continuum emission, using the relation:
\begin{equation}\label{equ:column}
M_{\rm gas} = \frac{\mathrm{g} S_{\nu} d^2}{B_{\nu}(T_{\rm dust}) \kappa_{\nu}},
\end{equation}
where $\mathrm{g}$ is the gas-to-dust mass ratio (assumed to be 100), $S_{\nu}$ is the flux density, $d$ is the source distance (8.3~kpc), $B_{\nu}(T_{\rm dust})$ is the Planck function at the dust temperature $T_{\rm dust}$, and $\kappa_{\nu}$ is the dust opacity. We adopted $\kappa_{\nu} = 1.99\ \mathrm{cm^2\ g^{-1}}$, corresponding to dust grains without ice mantles at a gas density of $10^6\ \mathrm{cm^{-3}}$ \citep{Ossenkopf1994}.
Given the high densities expected in accretion disks ($\gtrsim 10^7\ \mathrm{cm^{-3}}$), gas and dust are likely coupled effectively \citep[$T_{\rm gas} \approx T_{\rm dust}$;][]{Goldsmith2001}. We therefore adopted the derived CH$_3$OH rotational temperature as a proxy for the dust temperature ($T_{\rm dust} \approx T_{\rm ex}$).

The measured continuum flux density likely includes non-dust contributions, particularly free-free emission from the central high-mass protostar. However, we do not have high resolution radio continuum data to constrain the free-free emission directly from the protostar. To isolate the thermal dust emission from the disk, we modeled the central component as an unresolved point source by fitting a 2D Gaussian profile to the peak emission, with the FWHM fixed to the synthesized beam size. This central component was subsequently subtracted from the original map to mitigate contamination from the protostellar free-free or hot dust emission, thereby yielding a residual map that more accurately traces the dust distribution within the disk.

The disk mass is calculated on a pixel-by-pixel basis. For the area with a CH$_3$OH rotational temperature solution, we used the spatially resolved $T_{\rm ex}$, yielding a mass of 4.1~$M_{\odot}$. In addition, for pixels with continuum emission but lacking CH$_3$OH rotational temperatures, we adopted the mean temperature of $189$~K. The latter approach results in a total disk mass of 5.5~$M_{\odot}$. The 5.5~$\msol$ disk mass derived via \autoref{equ:column} assumes optically thin 1.3~mm emission ($\tau\ll1$). However, the high column densities expected around a $\sim$40~$\msol$ protostar inevitably lead to optically thick emission in the inner disk regions \citep{Fengwei2025}. Because this optical depth attenuates the observed flux, our 5.5~$\msol$ estimate represents a lower limit to the true disk mass.

\subsection{Modeling the molecular lines}\label{sec: model lines}

\autoref{fig:cont and moment1} presents the dust continuum structure and gas kinematics in the Sgr~C disk. The left panel displays the ALMA 1.3 mm continuum emission, revealing a compact dust structure associated with the high-mass protostar. The right panel shows the intensity-weighted mean velocity map derived from the stacked CH$_3$OCHO transitions (see \autoref{subsec: model lines}). This map reveals a rotating structure with velocities ranging from $\sim -56$\,km\,s$^{-1}$ (blue-shifted) to $\sim -46$\,km\,s$^{-1}$ (red-shifted). In \autoref{subsec:NMF}, we analyzed the spatial segregation of molecular species. In \autoref{subsec: model lines}, we employed a composite dynamical model to constrain the protostellar mass and disk structure. 

\subsubsection{NMF analysis of COMs}\label{subsec:NMF}

\begin{figure*}[!thpb]
\centering
\includegraphics[width=0.95\textwidth]{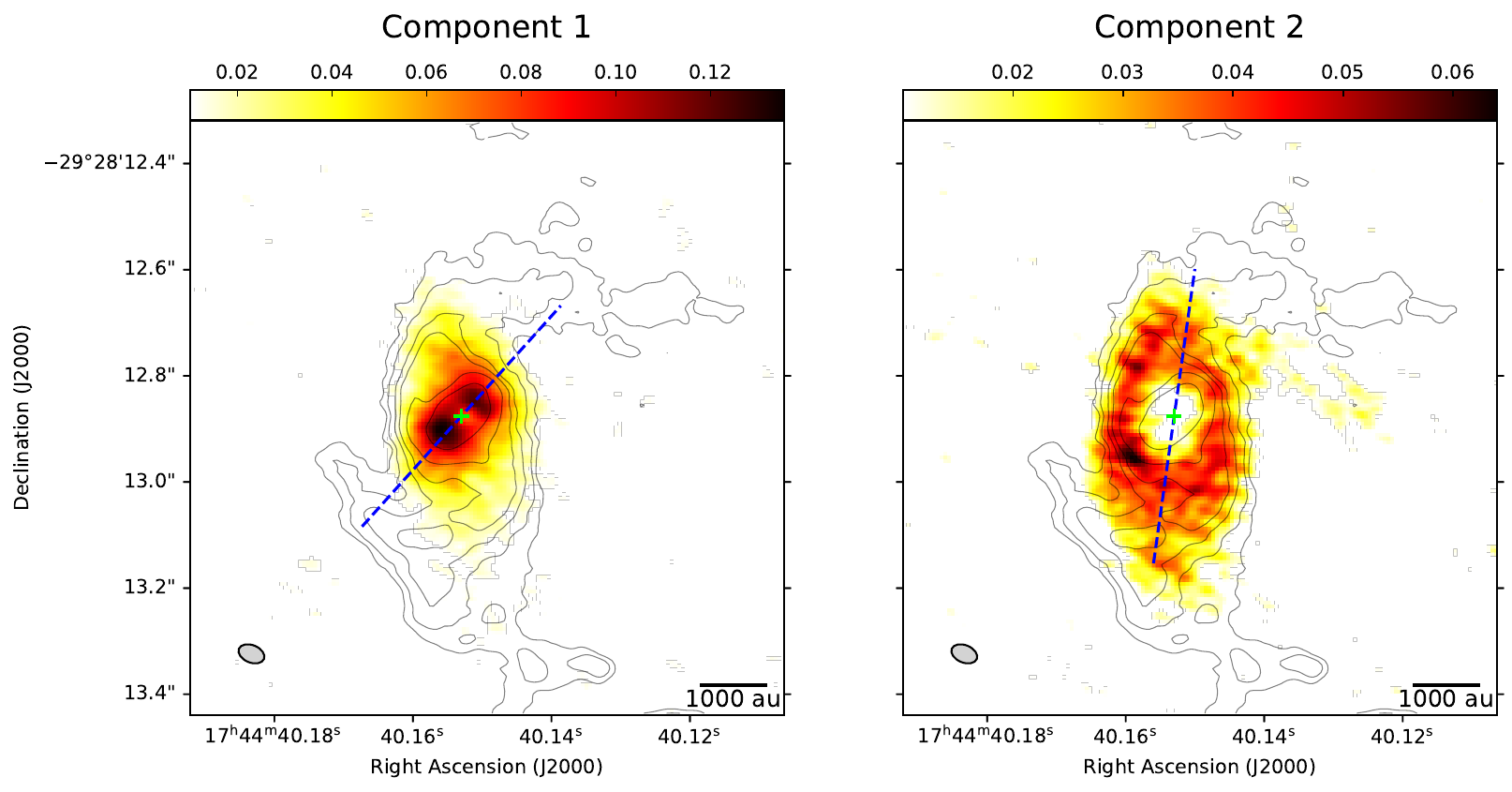}
\caption{Normalized intensity maps of the two NMF components, derived from the analysis of 40 molecular lines. The black contours represent the 1.3~mm dust continuum emission as described in \autoref{fig:cont and moment1}. The blue dashed line represents the visually estimated geometric principal axis, and the green cross represents the position of the protostar.
}
\label{fig:NMF}
\end{figure*}

Building upon the line identification presented in \autoref{subsec: line identification}, we generated velocity-integrated intensity maps for the 52 spectral lines listed in \autoref{tab:lines}, which are shown in \autoref{app:moment0}. A visual inspection of these maps suggests that the identified species reveal clear spatial differentiation across the disk. For instance, nitrogen-bearing species such as CH$_3$CN are concentrated in a central region, whereas oxygen-bearing species including CH$_3$OCHO and CH$_3$OH exhibit a more extended morphology.

To systematically analyze these varying spatial distributions, we employed a dimensionality reduction technique. Non-negative Matrix Factorization (NMF) has emerged as an effective method for extracting characteristic features from large astronomical datasets. NMF performs a low-rank approximation by decomposing the observed spectral-spatial matrix V into a set of non-negative basis spectra (H) and their corresponding spatial weights (W). Unlike Principal Component Analysis (PCA), which relies on orthogonal eigenvectors that can take negative values, NMF imposes a strict non-negativity constraint (H, W $\ge$ 0). This constraint is physically motivated by the additive nature of astronomical emission, preventing nonphysical negative fluxes and ensuring that each extracted component represents a distinct, localized physical environment rather than a mere statistical residual. Recent studies have successfully applied NMF to extragalactic multi-line data \citep{deMijolla2024, Kishikawa2025}, demonstrating its ability to decompose molecular emission originating from different physical components.

We applied NMF to the dataset of 40 molecular lines (see \autoref{tab:lines}) to isolate the principal components of molecular emission within the Sgr~C disk. We excluded 12 spectral lines from the NMF analysis, such as transitions of H$_2$CO, CS, and SO, because they exhibit a strong spatial asymmetry, showing virtually no emission in the southern part of the disk.

The NMF analysis extracts two main components with distinct spatial distributions, as shown in \autoref{fig:NMF}. The fractional contribution of each molecule line to the two components is detailed in \autoref{app:nmf_fractions}. We found a clear chemical segregation between the components: nitrogen-bearing molecules (e.g., CH$_3$CN, $^{13}$CH$_3$CN) mainly contribute to Component 1. In contrast, oxygen-bearing molecules (e.g., CH$_3$OH, CH$_3$OCHO) are found in both components, with $\mathrm{CH_3OCHO}$ being the primary contributor to Component 2. This implies the presence of two different physical and chemical environments within the Sgr~C disk, which then lead to the differential formation or excitation of these molecules \citep[e.g.,][]{Ginsburg2017,Csengeri2019,Qin2022,Taniguchi2023,Yonetsu2025}.

Component~1 is co-spatial with the inner elliptical dust continuum. It exhibits a central depression coincident with the protostellar position. Since NMF was performed on continuum-subtracted data, this depression is likely attributed to the attenuation of line emission by the high dust optical depth. At the location of the peak continuum emission, the derived CH$_3$OH rotational temperature is found to be 260~K. This temperature, when coupled with the observed dust emission intensity of $12.5~\mathrm{mJy\,beam^{-1}}$ at the same position, implies a high optical depth $\tau_v\approx2$. In such a regime, the line intensity measured after continuum subtraction is suppressed because the dust not only emits as a background but also absorbs the line photons originating from the protostar.

The extended distribution of O-bearing species in Component~2 suggests they may form early on grain surfaces and can be released into the gas phase even in the cooler envelope, while the compact distribution of N-bearing species implies that their liberation from ice mantles is triggered by the elevated temperatures found in the inner regions, where higher densities and radiation fields further drive their strong excitation \citep{Bisschop2007,Suzuki2018}. We posit that this observed chemical dichotomy implies the existence of two physically different structural components, possibly corresponding to an inner disk and an outer envelope region. Moreover, Component~2 displays a morphology whose main axis is offset in position angle by $\sim 25^{\circ}$ relative to Component~1. This position angle misalignment between the two components suggests the possible presence of a warped structure \citep{Kraus2020,Sai2020}.

\subsubsection{Fitting the PPV cube} \label{subsec: model lines}
Molecular line observations allow us to quantify the gas dynamical and thus to constrain the protostellar mass, disk structure, and infall rate. Previous studies suggest that dynamical modeling can be used to robustly constrain the dynamical parameters of the accretion disk without considering other physical mechanisms and structures of the accretion disk \citep{Maud2019,Zhang2019,Boyden2020, Chengyu2022}. 

The analysis in \autoref{subsec:NMF} revealed that the CH$_3$OCHO transitions trace both an inner, compact structure and a more extended, outer component. This spatial segregation and the channel maps of the CH$_3$OCHO transitions motivate our adoption of a composite dynamical model consisting of an inner Keplerian disk surrounded by an outer rotating and collapsing envelope. In line-rich regions, stacking multiple transitions effectively enhances the signal-to-noise ratio (S/N), which facilitates a more robust extraction of kinematic information \citep[e.g.,][]{Ginsburg2018,Schw2019}. Therefore, we stacked all CH$_3$OCHO transitions in \autoref{tab:lines} for subsequent dynamical modeling. We assigned a weight to each transition based on its integrated intensity. We performed a Gaussian fit to the spatially integrated spectrum of each identified CH$_3$OCHO line to derive its integrated flux, which were then used as weight factors during the stacking.

For the Keplerian disk component, we used the methodology of \citet{Chengyu2022}. The model assumes that the disk has a scale height relative to the midplane $h(r) = 0.2 \times r$ and is sharply truncated at an inner boundary $R_{\rm in}$ and an outer radius $R_{\rm c}$ where it joins the envelope. For the envelope, we adopted the Ulrich infall model \citep{Ulrich1976}, which provides an analytical solution for the collapse of spherically symmetric gas in uniform solid-body rotation when pressure forces are negligible. The material arriving at the mid-plane collides with material from the opposite $z$-direction, dissipating kinetic energy in shocks and forming a flat disk. The envelope extends to a radius $R_{\rm out}$.  For a detailed description of the model implementation, see \citet{Ulrich1976} and \citet{Chengyu2022}. We also performed fits using a standalone pure Keplerian disk or an Ulrich infall model, respectively. However, neither could adequately reproduce the observed kinematics. The composite dynamical model outperforms these individual components, providing a superior fit to the complex velocity structure of the Sgr~C disk.

As discussed in \autoref{subsec:NMF}, the two components in the NMF analysis show different position angles, suggesting the presence of a warped structure. We therefore adopted two different position angles (from North to West): $\theta_{\rm env}$ for the outer envelope and $\theta_{\rm disk}$ for the inner Keplerian disk. $M_{\star}$, the mass of the central protostar, was previously estimated to be $\sim 40\ M_{\odot}$ based on the luminosity \citep{LuXing2022}, which was used as the prior. Both the envelope and the disk are viewed at an inclination angle $i$, where $0^{\circ}$ corresponds to being face-on and $90^{\circ}$ being edge-on. To account for the density discontinuity between the inner and outer components, we added $m_{\rm env}/m_{\rm disk}$, the mass ratio between the envelope and Keplerian disk, to properly characterize the density contrast between the two components. The dynamic model does not constrain the absolute intensity. Therefore, we introduced a normalization factor, $f_{\rm norm}$, as a parameter to compare with observations.

We generated a synthetic Position-Position-Velocity (PPV) cube from this model. For each grid cell in the model, we first computed the line-of-sight velocity, $v_{\rm los}$. The line intensity at a specific velocity $v_{\rm obs}$ for any sky position is then obtained by integrating the contributions of all cells, weighted by density $\rho_i$ and the line emission profile $\phi(v_{\rm obs})$, along the line of sight. The line emission profile is described by
\begin{equation}
\phi(v_{\rm obs}) \propto \exp \left[ -\frac{(v_{\rm obs} - v_{\rm i,los})^2}{2\sigma^2} \right],
\end{equation}
where the thermal broadening line width $\sigma=\sqrt{kT/\mu \mathrm{m}_\text{H}}$, and $\mu$ is the molecular weight of the molecule in use. $T$ is the average temperature estimated in \autoref{subsec: temperature and mass}. This process yields a model PPV cube, which is then smoothed to match the spatial and spectral resolution of the observations.

We determined the best-fit parameters by comparing the model to the observation using a chi-squared ($\chi^2$) likelihood. For subsequent comparison in the PPV cube, both the best-fitting model and the data are normalized by their respective peak intensities. We employed the Markov Chain Monte Carlo (MCMC) sampler \textit{emcee} \citep{Foreman-Mackey2013} to efficiently explore the parameter space. The parameters explored in the fit are: $\theta_{\mathrm{env}}$, $\theta_{\mathrm{disk}}$, $M_{\star}$, $R_{\rm c}$, $R_{\rm in}$, $R_{\rm out}$, $i$, $m_\text{env}/m_\text{disk}$, and $f_\text{norm}$. \autoref{tab:initial parameters} shows the parameters and their priors in the model.

\begin{deluxetable*}{lccccc}
\tablecaption{Parameter priors and best-fit parameters\label{tab:initial parameters}}
\tablehead{
\colhead{Parameter} & \colhead{Priors}  & \colhead{Best-fit} & \colhead{Unit}
}
\setlength{\tabcolsep}{12pt}
\renewcommand{\arraystretch}{1.2}
\startdata
$\theta_{\mathrm{env}}$ & ($-$10, 60) & 18.2$^{+4}_{-6}$  & $\arcdeg$\\
$\theta_{\mathrm{disk}}$ & (0, 90)  & 42.6$^{+3}_{-5}$  & $\arcdeg$\\
$M_*$ & (20, 50) & 40$^{+2}_{-3}$  & $M_{\odot}$ \\
$R_{\mathrm{in}}$ & (80, 400) & 240$^{+50}_{-40}$  & au   \\
$R_{\mathrm{c}}$ & (400, 1600) & 1300$^{+100}_{-120}$  & au    \\
$R_{\mathrm{out}}$ & (1600, 2500)  & 1800$^{+80}_{-100}$  & au \\
$i$ & (30, 90)  & 50$^{+4}_{-4}$  & $\arcdeg$    \\
$m_\text{env}/m_\text{disk}$ & (0.5,2) & 1.1$^{+0.1}_{-0.1}$  &  \\
$f_\text{norm}$ & (0.7,1.3) & 1.08$^{+0.06}_{-0.04}$  &  \\
\enddata
\end{deluxetable*}
its LSR velocity is around −50 km s−1,
We used the observations to inform the priors. $v_{\rm sys}$ is the system velocity. The prior for $v_{\rm sys}$ is centered at $-51 \text{ km s}^{-1}$, based on the identification of molecular lines. The priors for the positional offset are centered at $\alpha$(J2000), $\delta$(J2000)= $17^{\rm h}44^{\rm m}40.153^{\rm s}$,  $-29\arcdeg 28\arcmin 12.876\arcsec$, corresponding to the peak of the dust continuum emission, which was assumed to be the location of the protostar. 

\begin{figure*}[!thpb]
\centering
\includegraphics[width=0.45\textwidth]{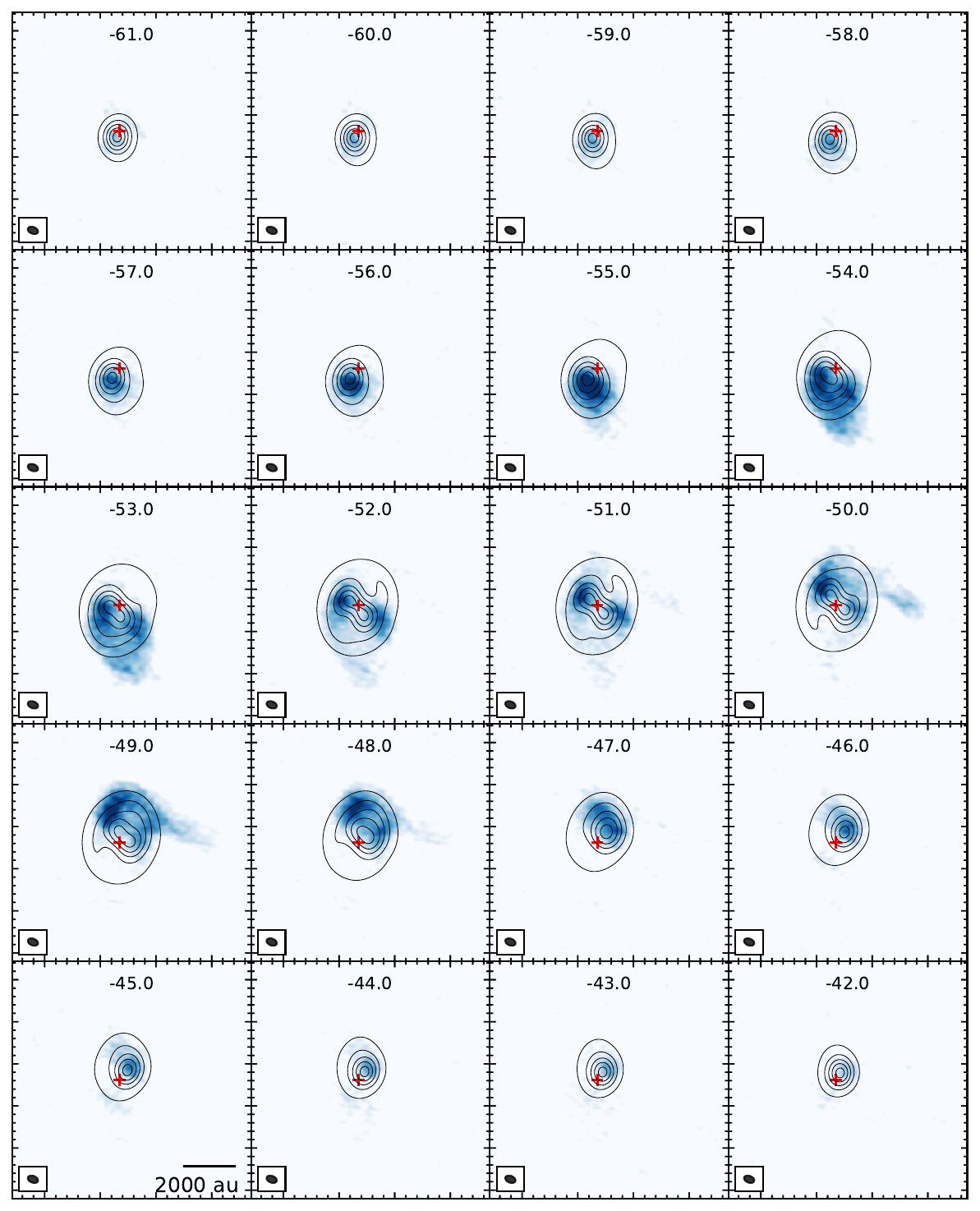}
\includegraphics[width=0.5\textwidth]{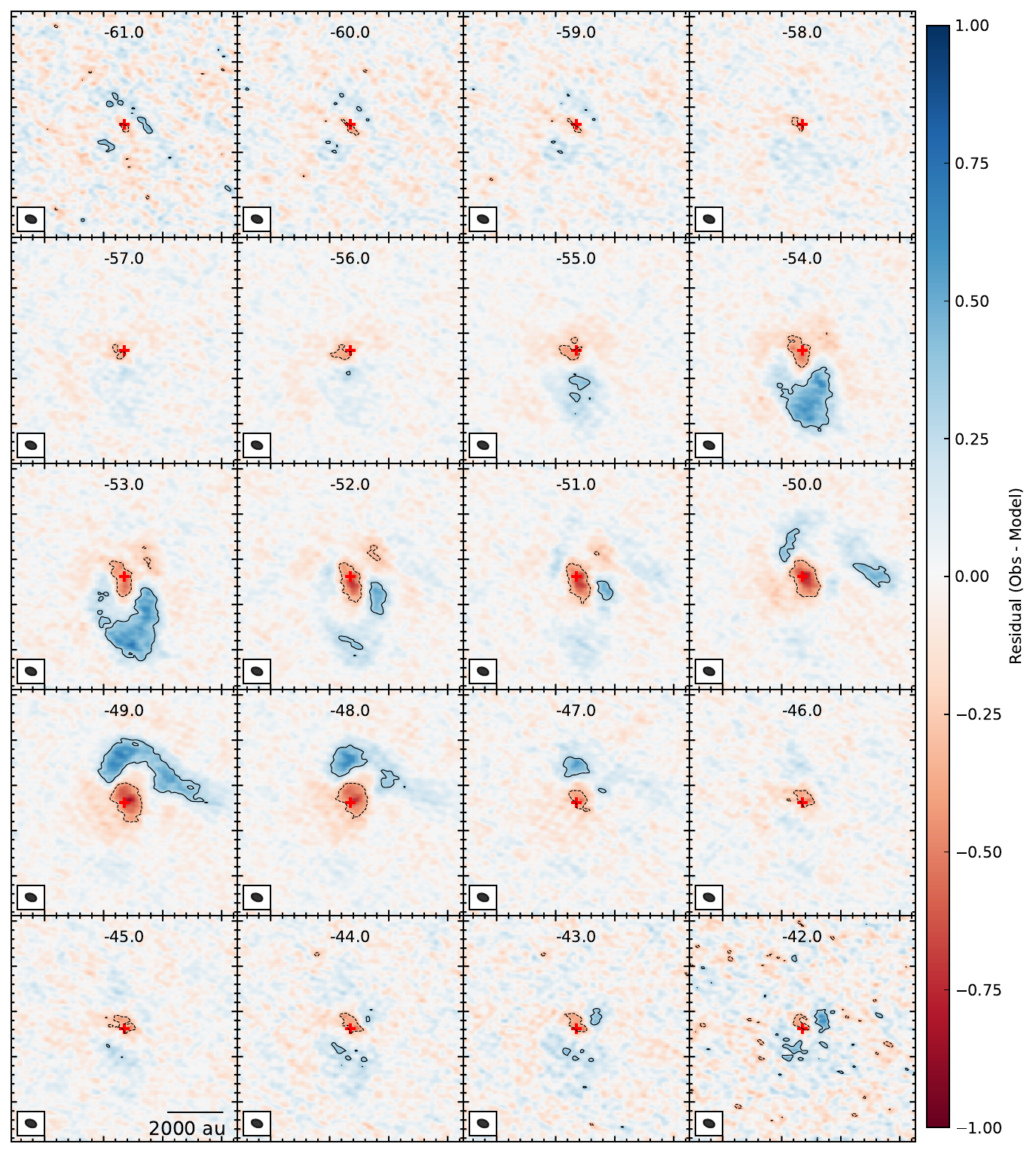}
\caption{
\textbf{Left:} Channel map (color scale) of the stacked CH$_3$OCHO. The LSR velocity in unit of km\,s$^{-1}$ of each channel is labeled in each panel. The black contours show the normalized intensities from the best-fit model, with levels at [0.1, 0.3, 0.5, 0.7, 0.9]. The red cross marks the position of the high-mass protostar.
\textbf{Right:} Residual map showing the difference between the normalized observed data and the model (Observed data$-$Model). The contour represents the value of 0.3.
}
\label{fig:channel map}
\end{figure*}

\autoref{fig:channel map} presents the results of dynamical modeling. Despite the localized residuals in the right panels, the model reproduces the large-scale velocity field of CH$_3$OCHO. For example, the observed high intensity emission far from the protostar in channels near the systemic velocity is reproduced by the model. This suggests that the structure of the Sgr~C disk can be described by the composite model: an inner Keplerian disk embedded in an outer, free-falling envelope. 

In the residual plot, the positive residual intensity is consistent with the position of the spirals that are noted in \citet{LuXing2022}. This is because our model does not include the spirals, whose dynamics will be discussed in \autoref{sec:streamer}. The negative residual is where the model is greater than the observed data, mainly distributed in the central part. The main reason is that the intensity of the CH$_3$OCHO lines that we selected is relatively weak in the interior (see \autoref{app:nmf_fractions}).

The best-fit parameters from this model are listed in \autoref{tab:initial parameters}. The fitting results show that the disk has a centrifugal radius at about 1300~au and an inner radius at 240~au, and the envelope has an outer radius at about 1800~au. The normalization factor, $f_{\rm norm}$, consistently converges to a value of $\approx 1.1$ in the fit. This indicates that the intensity scale of the model provides a good match to the observations, and normalization does not bias the determination of the dynamical parameters. We obtained an outer envelope to inner disk mass ratio of 1.1 from the best-fit results, and by substituting this into the total mass of 5.5~$M_{\odot}$ estimated in \autoref{subsec: temperature and mass}, we found the mass of the envelope to be 2.9~$M_{\odot}$. 

To further validate the composite dynamical structure derived from our 3D PPV dynamical modeling and to independently constrain the mass of the central protostar, we conducted a position-velocity (PV) analysis along the major axis of the Sgr~C disk. By extracting the velocity structures at different signal-to-noise thresholds, we can isolate the rotation curves of the inner and outer regions. The PV fitting reveals a purely Keplerian rotation profile ($v_\text{rot}\propto r^{-0.5}$) within $1300~\mathrm{au}$, and a steeper profile ($v_\text{rot}\propto r^{-0.8}$) extending out to $1800~\mathrm{au}$ that is characteristic of an infalling envelope. The detailed fitting of the PV diagram, which yields a central stellar mass of $M_* \sim 40~M_{\odot}$, is presented in \autoref{app: PV}.

\subsection{Estimating the accretion rate}\label{subsec: accretion rate}
Directly measuring the accretion rate $\dot{M}_{\rm acc}$ for deeply embedded high-mass protostars is challenging. We therefore turn to indirect methods that probe the infall rate of material from the larger-scale envelope to disk, $\dot{M}_{\rm inf}$. While these techniques measure the mass supply rather than the instantaneous accretion onto the protostar itself, they provide crucial constraints on the mass flow in these obscured systems. In \autoref{subsec:NMF}, we analyzed the emission intensity and velocity field of the spectral lines in the Sgr~C disk, and applied the results to the dynamical model in \autoref{subsec: model lines}. Then we estimated the accretion rate based on the results of the model fitting.

The accretion rate $\dot{M}$ is derived by linking the total envelope mass $M_{\text{env}}$ to the local density $\rho_1(r, \theta)$ predicted by the Ulrich (1976) model. The total mass within a given volume $V$ is the integral of the density over that volume:
\begin{equation}
    M_{\text{env}} = \int_{V} \rho_1 dV.
\end{equation}
The density of the model is directly proportional to $\dot{M}$, allowing the accretion rate to be factored out of the integral:
\begin{equation}
    M_{\text{env}} = \dot{M} \int_{V} \left( \frac{\rho_1}{\dot{M}} \right) dV.
\end{equation}  
This allows $\dot{M}$ to be solved directly:
\begin{equation}
    \dot{M} = \frac{M_{\text{env}}}{I},
    \label{eq:mdot_solution}
\end{equation}
where $I$ is the integral of the normalized density distribution, which we computed numerically:
\begin{equation}
    I = \int_{V} \left( \frac{\rho_1}{\dot{M}} \right) dV.
    \label{eq:I_integral}
\end{equation}
The normalized density $(\rho_1 / \dot{M})$ is derived from the density equation of the model~\citep{Ulrich1976}:
\begin{equation}\label{eq:rho_norm}
\begin{split}
    \frac{\rho_1}{\dot{M}} = & \frac{1}{8\pi(GMr^3)^{1/2}} \left(1 + \frac{\cos\theta}{\cos\theta_0}\right)^{-1/2} \\
    & \times \left(\frac{\cos\theta}{2\cos\theta_0} + \frac{r_d}{r}\cos^2\theta_0\right)^{-1}.
\end{split}
\end{equation}
The primary challenge in computing this integral is that the variable $\theta_0$ (the initial angle of the streamline at infinity) is implicitly defined at every point $(r, \theta)$ by the streamline equation:
\begin{equation}
    r = \frac{r_d \cos\theta_0 \sin^2\theta_0}{\cos\theta_0 - \cos\theta},
    \label{eq:streamline}
\end{equation}
where $r_d$ is the centrifugal radius. To evaluate the integral $I$, we first rearranged \autoref{eq:streamline} into a cubic polynomial for $x = \cos\theta_0$:
\begin{equation}
    r_d x^3 + (r - r_d)x - r\cos\theta = 0.
    \label{eq:cubic}
\end{equation}
At each integration step for a given $(r, \theta)$, we first numerically solved \autoref{eq:cubic} to find the corresponding $\cos\theta_0$. We then performed a 2D numerical integration over $r \in [r_{\text{in}}, r_{\text{out}}]$ and $\theta \in [0, \pi]$.  This value is then substituted into \autoref{eq:rho_norm} to evaluate the integrand. The resulting value of the integral $I$ is then used in \autoref{eq:mdot_solution} to find the accretion rate $\dot{M}$. By adopting the best-fit results from \autoref{subsec: model lines}, we obtained $\dot{M}_{\rm inf} = 7 \times 10^{-3}\ M_{\odot}\ \mathrm{yr^{-1}}$. Considering the uncertainties in the estimate, such as the use of an idealized symmetric model, the fitting errors of \textit{emcee}, and the error in the disk mass, the result should be considered as an order-of-magnitude estimate.

\section{Discussion} \label{sec:discussion}

\subsection{Spirals in the Sgr~C disk as infalling streamers} \label{sec:streamer}

\begin{figure*}[!thpb]
\centering
\includegraphics[width=0.4\textwidth]{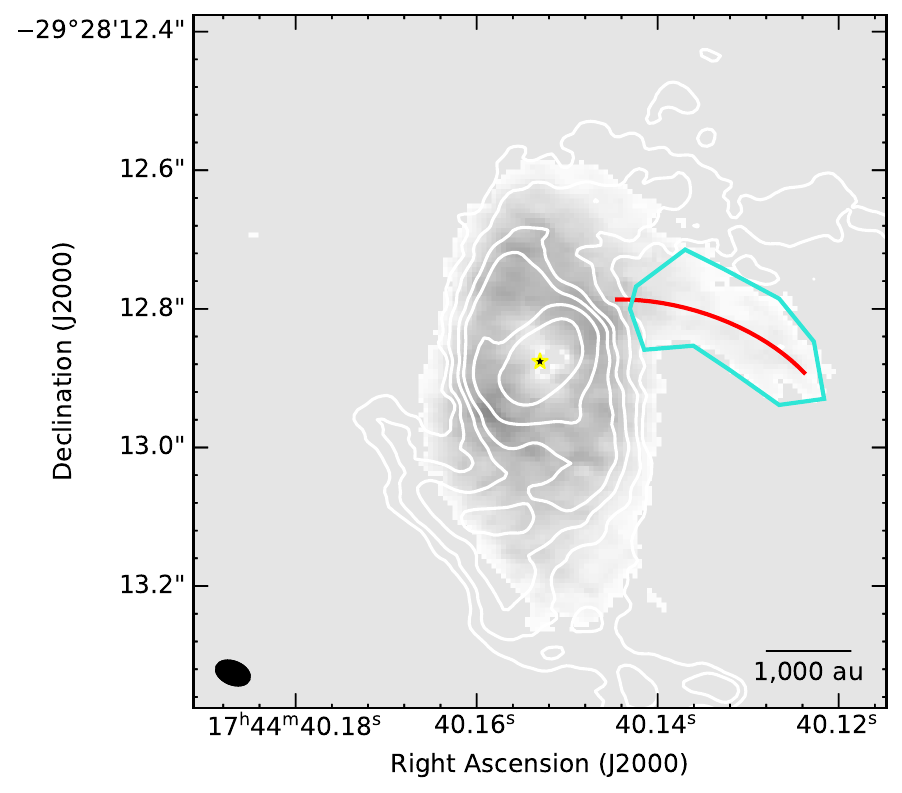}
\includegraphics[width=0.43\textwidth]{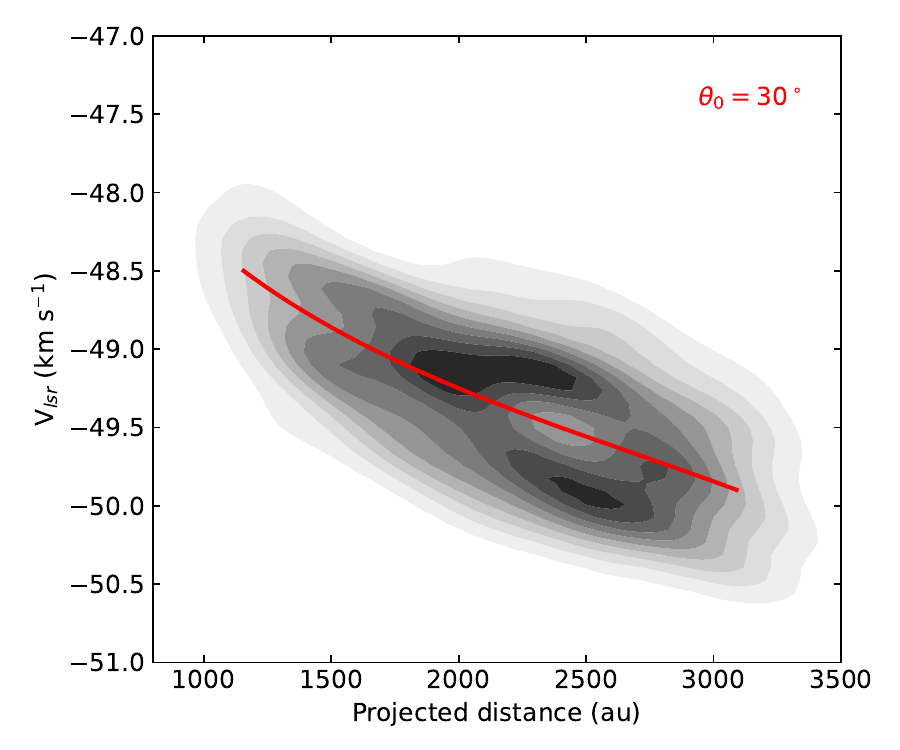}
\includegraphics[width=0.4\textwidth]{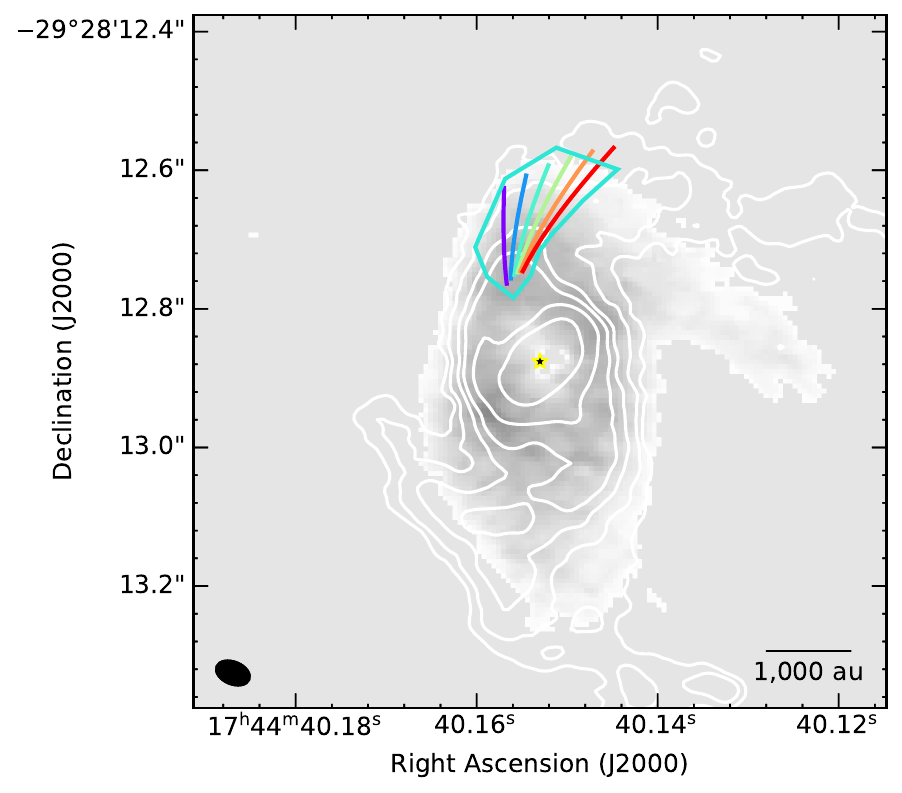}
\includegraphics[width=0.43\textwidth]{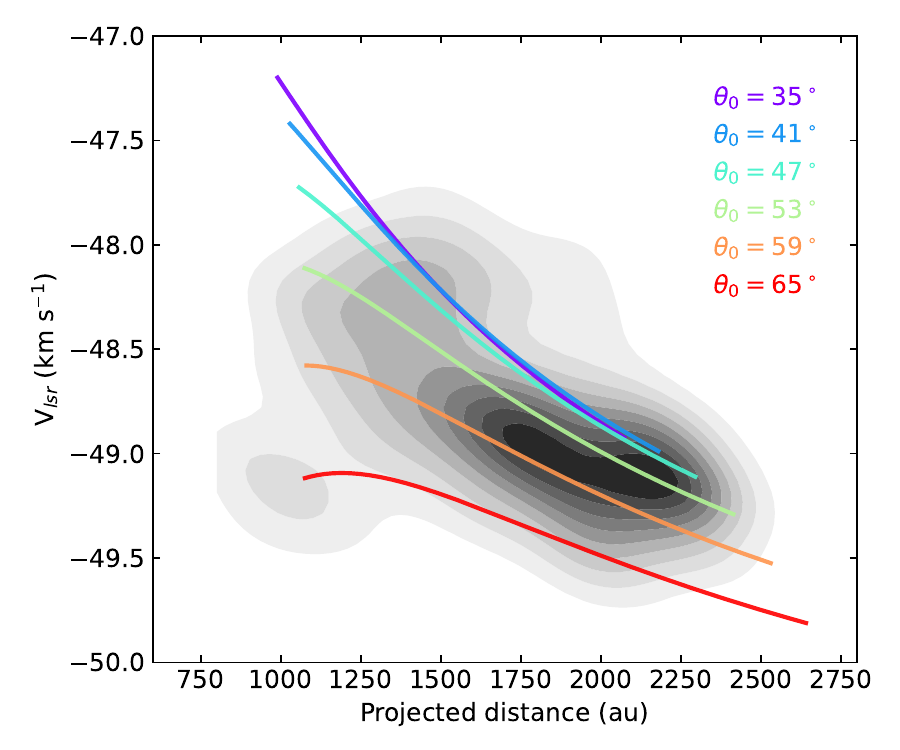}
\includegraphics[width=0.4\textwidth]{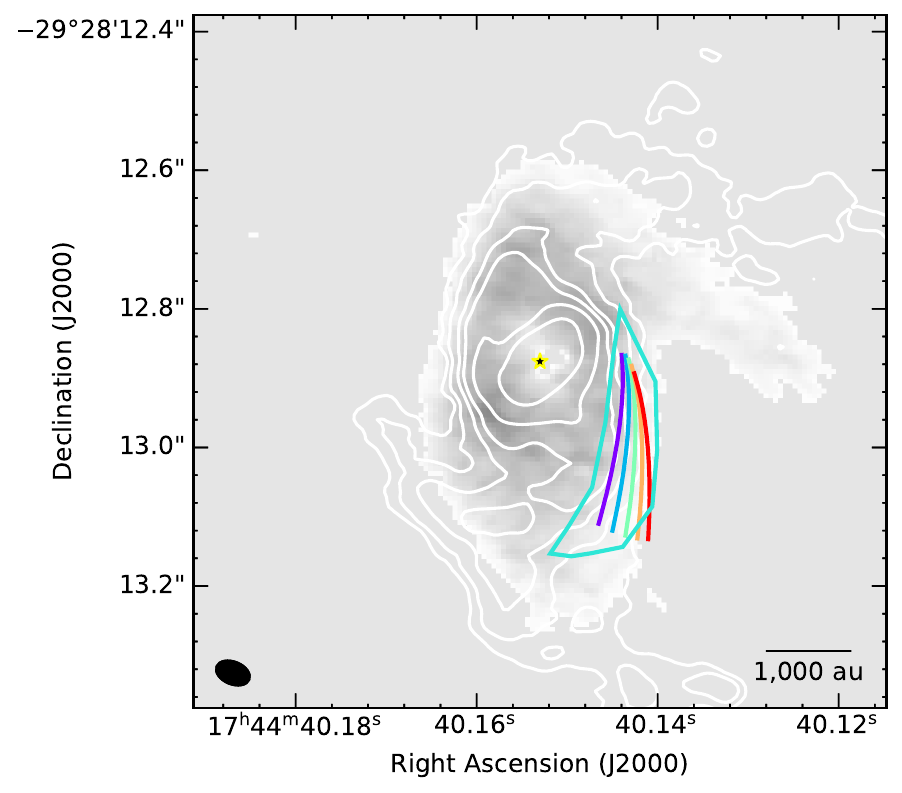}
\includegraphics[width=0.43\textwidth]{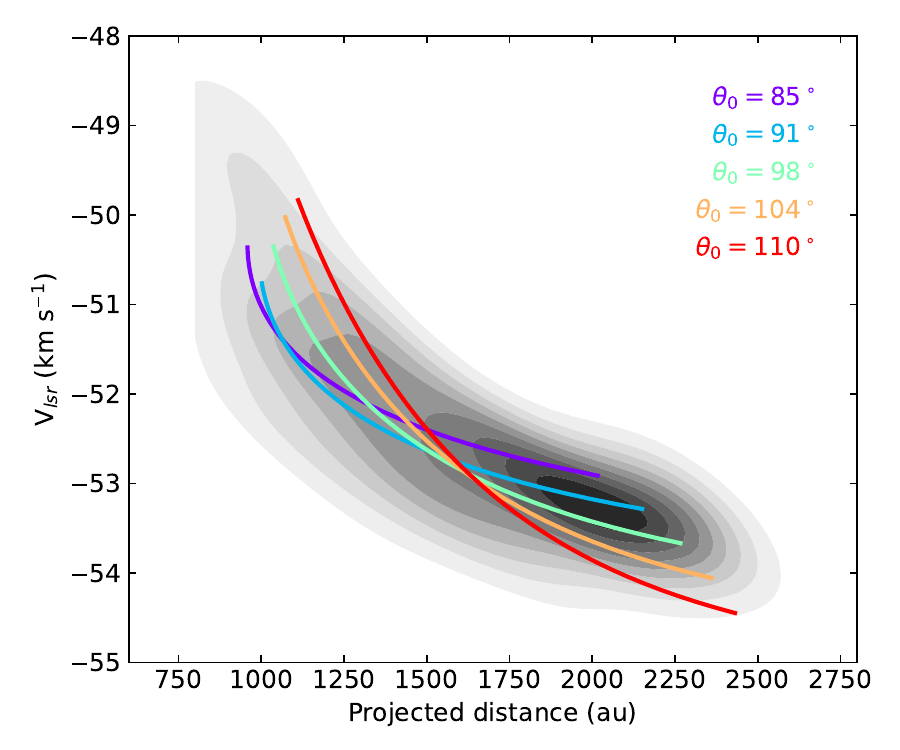}
\caption{Intensity maps of stacked CH$_3$OCHO emission overlaid with model trajectories of infalling and rotating particles. The white contours denote the continuum emission as described in \autoref{fig:cont and moment1}. Different colors represent particle trajectories originating from different polar angles. The yellow star represents the position of the protostar. The three rows correspond to the West spiral (Ws, top), North spiral (Ns, middle), and South spiral (Ss, bottom), respectively. The two columns display the peak intensity maps of stacked CH$_3$OCHO (left) and the kernel density distributions of projected distance versus centroid velocity (right).}
\label{fig:streamer_line}
\end{figure*}

Although the overall kinematics in the outer part of the Sgr~C disk are well described by the axisymmetric Ulrich model, the observed spiral structures reveal additional accretion pathways, likely feeding the inner disk via streamers. As shown in \autoref{fig:cont and moment1}, the continuum emission of the Sgr~C disk exhibits two spirals located in the north and south. In contrast, the velocity field of the stacked CH$_3$OCHO displays kinematic structures that differ from the continuum morphology, most notably revealing an extended spiral feature in the west out to $3500$~au (see \autoref{fig:streamer_line} top). The spatial distribution of the northern and southern spirals in the continuum aligns with the envelope structure seen in the centroid velocity map. We refer to these structures as the West spiral (Ws), North spiral (Ns), and South spiral (Ss), respectively, based on their positions with respect to the central protostar. In \autoref{fig:streamer_line}, the cyan contours correspond to the areas of the positive intensities in the residual map of \autoref{fig:channel map}, where the three spirals are prominent. 

To investigate the dynamical properties of the spiral features, we model the kinematics observed in the stacked CH$_3$OCHO emission using analytic solutions for a particle infalling toward a central mass within rotating gas. This approach allows us to test whether the observed velocity gradients are consistent with accretion flows or streamers. The velocity distribution is approximated by that of a rotating and collapsing envelope with conserved angular momentum \citep{Mendoza2009}, defined by the central mass $M_{*}$, initial radius $r_0$, centrifugal radius $r_{c}$, initial polar angle $\theta_0$, initial azimuthal angle $\phi_0$, and initial radial velocity $v_{0}$. 

Adopting the system parameters derived in \autoref{subsec: model lines}, we fix the central mass at $M_{*}=40~M_{\odot}$ and the position angle at $42^{\circ}$. The position angle is measured clockwise from the North, and the polar angle is defined relative to the spin axis of a plane with an inclination angle of $50^{\circ}$. We tune the free parameters to reproduce the observed kinematic features. For Ws (\autoref{fig:streamer_line} top), the best-match model parameters are $r_0=2400$~au, $r_{cb}=1200$~au, $\theta_0=30^{\circ}$, $\phi_0=95^{\circ}$, and $v_0=1.5~\mathrm{km\ s^{-1}}$. The kinematics of Ns and Ss are best explained by a family of trajectories with varying initial polar angles (\autoref{fig:streamer_line} middle and bottom). The adopted parameters are: $r_0=3000$~au, $r_{cb}=1100$~au, $\theta_0=[35^{\circ},41^{\circ},47^{\circ},53^{\circ},59^{\circ},65^{\circ}]$, $\phi_0=-15^{\circ}$, and $v_0= 6\ \mathrm{km\ s^{-1}}$ for Ns; and $r_0=2500$~au, $r_{cb}=1200$~au, $\theta_0=[85^{\circ},91^{\circ},98^{\circ},104^{\circ},110^{\circ}]$, $\phi_0=-125^{\circ}$, and $v_0= 4.5\ \mathrm{km\ s^{-1}}$ for Ss.

These streamline models reproduce the observations, supporting the interpretation as streamers undergoing gravitational collapse and rotation toward the central protostar. Similar streamer structures have been identified in other massive star formation regions, where asymmetric accretion flows are ubiquitous \citep{Chen2020,Burns2023,Olguin2025,Xiaofeng2025}. These streamers represent a non-axisymmetric mode of mass accretion that is likely critical for the rapid growth of high-mass protostars, which may deliver material directly from the envelope to the disk scales. Moreover, these streamers suggest that high-mass protostellar growth is not a purely localized process, but is coupled to the dynamics of the parent dense core.

Beyond accretion streamers, alternative mechanisms may account for the spiral structures observed in the Sgr~C disk. Morphologically, a visual inspection of the continuum emission reveals a potential spiral-like extension of Ss to the southeast. We do not model this feature due to the lack of corresponding molecular line emission. Nevertheless, this feature could represent a counterpart to the northern spiral. The overall geometry of Ns, Ss, and their potential extensions visually resembles an $m=2$ spiral density wave. Such $m=2$ modes are typically driven either by gravitational instability (GI)  \citep{Kratter2016,Meyer2018}. However, the Sgr~C disk maintains a sufficiently high Toomre stability parameter \citep[$Q\gg2$;][]{LuXing2022}, indicating that the substantial thermal and turbulent pressure in this warm environment could effectively suppress GI \citep{Toomre1964}. The $m=2$ modes could also be driven by tidal interactions from an unseen binary companion  \citep{Dong2015}, yet no companion has been resolved at the current observational resolution. Furthermore, recent theory suggests that anisotropic tidal forces from the central object can drive the formation of spiral accretion streamers \citep{GuangXing2024}. In summary, the streamer interpretation remains to be a physically coherent scenario, though future higher-resolution observations are required to more clearly elucidate the origin of these spiral structures.

\subsection{The evolution and fate of the high-mass protostar}
The stellar mass derived from our composite dynamical model places the protostar in early O-type. With a derived central mass of $M_* \approx 40~M_{\odot}$, the protostar has likely already engaged hydrogen fusion before reaching the zero-age main sequence (ZAMS) while remaining deeply embedded in its parent core. Our detection of a rotationally supported disk with a centrifugal radius at $R_{c} \approx 1300$~au provides direct observational evidence for the disk-mediated accretion scenario. The accretion rate $\dot{M} \approx 7 \times 10^{-3}~M_{\odot}~\mathrm{yr^{-1}}$ is similar to those found in high-mass sources in the Galactic disk, indicating that the suppression of star formation efficiency is scale-dependent in CMZ \citep[e.g.,][]{Kruijssen2014,LuXing2019b,Battersby2025}. It does not operate at the accretion disk scale dominated by gravity, supporting that evolution and star-forming activities of self-gravitationally bound gas structures may be self regulated, insensitive to the exterior environment on the full CMZ scale \citep{Linjing2026}. 

Although the total gas mass in the system is estimated to be $\sim 5.5\ M_{\odot}$, which would imply a rapid depletion timescale of $t_{\mathrm{dep}} \sim M_{\mathrm{env}}/\dot{M} < 1000$ years, large-scale gas dynamics suggest a sustained accretion process. The SMA observations of the $\sim 0.1$-pc dense core around the envelope reveal a virial parameter smaller than 1, suggesting that the core is gravitationally bound and acts as a gas reservoir of 500~$M_{\odot}$ to feed the envelope \citep{LuXing2019b}. To quantify this sustained feeding, we estimate the mass infall rate from this larger reservoir. Assuming a simple free-fall collapse of the 500~$M_{\odot}$ core from a radius of $R=0.1$~pc onto the $M_{*}=40\ M_{\odot}$ central object, the free-fall time is $t_\text{ff}=\frac{\pi}{2}\sqrt{R^3/(2GM_{\rm core})}=2.6\times 10^4$~yr. This yields a mean mass infall rate of $\dot{M}_\text{inf} = M_\text{core}/t_\text{ff} = 2.2 \times10^{-2}~M_{\odot}\,\mathrm{yr^{-1}}$. This infall rate is within the required magnitude to continuously replenish the inner envelope against its rapid depletion, indicating that the system is not transiently depleted but rather in a steady state that is consistent with an Ulrich-type model. In this picture, the $\sim0.1$-pc dense core provides the persistent mass budget, the envelope and streamers act as intermediate structures, and the gas is actively accreted into the central protostar through the inner disk.

We perform an order-of-magnitude estimate of the forces exerted by radiation from the central protostar and the infall of the envelope. Following \citet{Keto2006}, assuming that the radiation field of the protostar is spherically symmetric and that all stellar radiation is absorbed by dust, we estimate the magnitude of the radiation force $L/4\pi c$, where $L$ is the luminosity and $c$ is the speed of light. In comparison, the force exerted by the infall in a spherical accretion scenario is approximately $\dot{M}v_{\rm in}/4\pi$, where $v_{\rm in}$ is the infall velocity. Adopting a luminosity $L = 5.34 \times 10^5\ L_\odot$ , which is primarily dominated by a star with 40~$M_{\odot}$ \citep{Davies2011}, the magnitude of the outward force from the source radiation field is approximately $10^{27}$~dyn. On the other hand, the infall velocity at the centrifugal radius of $\sim$1300~au is $\sim 4~\mathrm{km\ s^{-1}}$ and the infall rate is $\dot{M}_{\rm inf} = 7 \times 10^{-3}~M_{\odot}\,\mathrm{yr^{-1}}$. Consequently, the magnitude of the inward force generated by the infalling envelope is approximately $10^{28}$~dyn. The lower limit of the force exerted by the streamers can be estimated by $M\dot{v}$, where $M=0.34~M_{\odot}$ for Ns, and $M= 0.37~M_{\odot}$ for Ss. The mass of Ws cannot be estimated, because it does not have available temperature or continuum emission measurement. We measure the average $\dot{v}$ from the streamlines in \autoref{fig:streamer_line}. Thus, the inward force exerted by the inflow in Ns and Ss is of the order of $10^{28}$~dyn. Therefore, the inward force of the infalling envelope or streamers is sufficient to overcome the outward radiation pressure from the protostar, even without considering the shielding effect of the disk \citep{Yorke2002,Krumholz2009}. Consequently, the gas is expected to ultimately be accreted onto the central protostar.

\subsection{Comparison with other accreting disks in the Galaxy}
\label{sec:chemistry}

\subsubsection{Chemical dichotomy in protostellar disks}
Our observations reveal a chemical dichotomy within the Sgr~C disk system: N-bearing COMs (e.g., CH$_3$CN) exhibit compact morphologies concentrated toward the inner Keplerian disk, whereas O-bearing species (e.g., CH$_3$OCHO and CH$_3$OH) present substantially more extended distributions (see \autoref{subsec:NMF}). Similar N/O dichotomies have been reported in high-mass protostellar disks outside the CMZ, such as NGC~6334~I(N), G11.92$-$0.61~MM1, and AFGL~4176 \citep{Hunter2014, Ilee2018, Johnston2020b}. In these sources, N-bearing species are typically found to be closely associated with the compact hot core emission, while O-bearing species display a more widespread distribution.

This chemical segregation extends to lower-mass systems as well. High angular resolution ALMA studies of Class~0/I sources frequently demonstrate that N-bearing COM emission is centrally peaked at small radii. In contrast, O-bearing species such as CH$_3$OH along with certain S-bearing shock tracers 
often exhibit ring-like structures peaking near the centrifugal barrier or along the disk-envelope interface \citep{Sakai2014, Oya2016, Artur2019}. Their extended morphologies are commonly interpreted as signatures of localized accretion shocks and subsequent nonthermal desorption, rather than thermal sublimation from the quiescent disk midplane \citep{Aota2015,Miura2017}.

To further investigate whether the extracted NMF components represent true chemical variations rather than radiative transfer effects in the Sgr~C disk, we analyze the upper-state energy ($E_{\rm u}$) distribution of the transitions associated with each component in \autoref{subsec:NMF}. The $E_{\rm u}$ distributions of the transitions constituting Component~1 and Component~2 exhibit substantial overlap, with both encompassing a broad range of excitation energies. The mean difference ($\langle\Delta E_{\rm u}\rangle$) is 19 K with a Welch t-test ($\rm p = 0.45$) confirm that there is no significant difference. The lack of a clear excitation gradient supports the explanation that the NMF decomposition reflects not a mere radiative transfer effect, but primarily a spatial dichotomy in chemical composition. Several mechanisms can account for this observed segregation. From an astrochemical perspective, \citet{Suzuki2018} hypothesized that N-bearing molecules are preferentially enhanced via gas-phase reactions in hot ($>$100~K) environments following the sublimation of icy dust mantles, as the hydrogenation of these species is inefficient at elevated temperatures. In contrast, the apparent central deficit of O-bearing COMs may reflect true inner-disk depletion caused by thermal processing or UV photodissociation.

However, this chemical interpretation must be treated with caution, as it is fundamentally entangled with radiative transfer effects. High dust continuum optical depth in the inner disk can suppress the line emission of certain extended species \citep{Galvan2018, Ginsburg2018, DeSimone2020, vanGelder2022}. Breaking the degeneracy between intrinsic chemical depletion and dust opacity will require future high-resolution, multi-band observations, especially those at lower frequencies where the continuum is optically thinner, to properly model continuum optical depths and excitation conditions.

In summary, unlike the situation on the cloud scale of a few pc in the CMZ, where the chemistry is different from that in nearby clouds \citep[e.g., widespread COMs in the CMZ that are driven by shocks;][]{MartinPintado1997,Requena2008,Menten2009}, we do not find evidence suggesting that the physics and chemistry governing COM distributions in the Sgr C disk are different from those in typical Galactic disk systems. 

\subsubsection{Accretion rate}
The estimation of a dynamical mass $M_{*} \approx 40~M_{\odot}$ and an infall rate $\dot{M}_\text{inf} \approx 7 \times 10^{-3}~M_{\odot}\,\text{yr}^{-1}$ positions the Sgr~C protostar at the high end of the mass spectrum among the known Galactic massive protostellar population. To place these properties in a broader context, we compare the Sgr~C system with a representative sample of well-characterized high-mass protostars hosting (sub\nobreakdash-)Keplerian disks or other rotating structures. 

The infall rate of the Sgr C protostar is similar to that of the high-mass protostar GGD27-MM1 ($M_{*} \sim 20~M_{\odot}$), which was estimated to be $5.2 \times 10^{-3}$--$1.6 \times 10^{-2}~M_{\odot}\,\text{yr}^{-1}$ via redshifted absorption profiles against the central continuum \citep{Fernandez-Lopez2023}. Similarly, in the gravitationally unstable disk of G353.273+0.641 ($M_{*} \sim 10~M_{\odot}$), \citet{Motogi2019} estimated a comparable envelope infall rate of $\ge 3 \times 10^{-3}~M_{\odot}\,\text{yr}^{-1}$ based on the envelope mass and free-fall timescale. The similar infall rates between the Sgr~C protostar and these sources, both derived from gas dynamics, suggests that the Sgr~C protostar is in an early, envelope-dominated phase where the mass supply from the parent core remains vigorous. The envelope ($\sim 10^3$--$10^4$~au) is actively feeding the inner disk via steady-state infall, sustaining a high-mass growth mode prior to the dispersal of the envelope.

Meanwhile, accretion rates for some disks are inferred directly from sub-Keplerian velocity profiles. For instance, detailed dynamical modeling of Keplerian disks in IRAS~20126+4104 ($M_{*} \sim 12~M_{\odot}$; \citealt{Cesaroni2025}) and Cepheus A HW2 ($M_{*} \sim 16~M_{\odot}$; \citealt{Sanna2025}) yields accretion rates of $\sim 1.7 \times 10^{-3}$ and $2.0 \times 10^{-3}~M_{\odot}\,\text{yr}^{-1}$, respectively. A similar range of $1.0$--$4.3 \times 10^{-3}~M_{\odot}\,\text{yr}^{-1}$ was derived for the W51 cluster members ($M_{*} \ge 20~M_{\odot}$) by indirectly converting outflow momentum fluxes to accretion rates \citep{Goddi2020}. The accretion rates derived from disk-to-protostar models are generally lower than the infall rates obtained from the envelope. This is anticipated due to mass loss through processes such as outflows \citep{JCtan2014}. Therefore, our derived infall rate serves as an upper limit for the final accretion rate onto the protostar.

The derived disk-to-star mass ratio $\mu = M_\text{disk}/M_{*} \approx 0.14$ places the disk near the regime where self-gravity becomes non-negligible \citep[$\mu > 0.1$;][]{Kratter2016}. In massive star-forming regions, such a high disk mass would trigger gravitational instabilities, leading to fragmentation or transient massive toroids, as observed in sources such as G31.41+0.31 \citep{Beltran2022}. However, the Sgr~C disk maintains sufficiently high Toomre $Q$ values \citep[$Q\gg2$;][]{LuXing2022}, indicating that the high thermal pressure in this warm environment effectively suppresses gravitational instabilities \citep{Toomre1964}. The Sgr~C system appears to host a stable, settled Keplerian disk that is efficiently fed by the envelope and streamers, as discussed in \autoref{sec:streamer}. A gravitationally stable inner disk embedded within and fed by a massive infalling envelope and streamers represents a robust pathway to sustain the high accretion rates required for early-O star formation.

\section{Summary}\label{sec:summary}
In this work, we present high-resolution ALMA Band 6（1.3~mm）molecular line observations of a massive protostellar disk candidate in the Sgr~C cloud of the CMZ, achieving a spatial resolution of 0\farcs{04} ($\sim$300~au at 8.3~kpc). By combining chemical analysis with dynamical modeling of the molecular lines, we characterize the structure, dynamics, and accretion properties of this early O-type protostellar system. Our main conclusions are summarized as follows.

\begin{enumerate}
    \item We derive the rotational temperature map using multiple transitions of $\mathrm{CH_3OH}$. The disk exhibits temperatures ranging from $\sim 100$~K to $\sim260$~K, with an average temperature of $189$~K derived from the rotational diagram. These high temperatures characterize a warm and dense environment in the accretion disk.
    
    \item Using Non-negative Matrix Factorization (NMF) on 40 molecular lines, we identify two different components governing the chemical distribution. Nitrogen-bearing species (e.g., $\mathrm{CH_3CN}$) are concentrated in the inner compact region, while oxygen-bearing species (e.g., $\mathrm{CH_3OCHO}$) trace a more extended structure. This chemical segregation coincides with the physical transition from the infalling envelope to the inner disk.
    
    \item We reproduce the kinematics of the stacked $\mathrm{CH_3OCHO}$ emission using a composite dynamical model consisting of an inner Keplerian disk and an outer free-falling (Ulrich) envelope. The best-fitting model constrains the central protostellar mass to $M_{\star} \approx 40~M_{\odot}$, and reveals a centrifugal radius at $R_{c} \approx 1300$~au that separates the rotationally supported disk from the infalling envelope. The fitting results also show that the disk has a inner radius at 240~au and the envelope has an outer radius at about 1800~au.
    
    \item Based on the envelope mass and density profile derived from the dynamical model, we estimate a mass infall rate of $\dot{M}_\text{inf} \approx 7 \times 10^{-3}~M_{\odot}~\mathrm{yr}^{-1}$. This rate is similar to typical values for protostars in the Galactic disk, suggesting a coherent picture of disk-mediated accretion around high-mass protostars in both the CMZ and the Galactic disk environments.
    
    \item We identify spiral features in both the dust continuum and molecular line velocity fields. Trajectory modeling confirms that these structures are consistent with being accretion streamers undergoing gravitational collapse toward the central protostar. These streamers likely serve as the mechanism replenishing the disk/envelope system.
\end{enumerate}
 
In conclusion, our study provides direct observational evidence for disk-mediated accretion onto an early O-type protostar ($\sim$40~$M_{\odot}$) within the extreme environment of the CMZ. The Sgr~C disk presents a scenario where a Keplerian disk, fed by an infall envelope and streamers, facilitates the formation of a high-mass star through rapid mass inflow. The operation of such a classical accretion mechanism in Sgr~C implies that while the extreme CMZ environment may suppress the global star formation efficiency at large scales, the accretion disk scale dominated by gravity is less affected by the environment. This indicates a universal disk-mediated accretion framework for high massive protostars across different environments of the Galaxy.

\begin{acknowledgments}
We thank Dr.~Yishen Qiu for his help on Spectuner. 
We thank Prof.~Siyi Feng for holding the Summer School of Astrochemistry in Xiamen in 2025, which provided a great platform of learning astrochemistry for the first author.
X.L.\ acknowledges support from the Strategic Priority Research Program of the Chinese Academy of Sciences (CAS) Grant No.\ XDB0800300, the National SKA Program of China (2025SKA0140100), the National Natural Science Foundation of China (NSFC) through grant Nos.\ 12273090 and 12322305, and the National Key R$\&$D Program of China (No.\ 2022YFA1603101), and the Natural Science Foundation of Shanghai (No.\ 23ZR1482100).
S.L.\ acknowledges support from the National SKA Program of China with No.\ 2025SKA0140100, ``Double First-Class'' Funding with No.\ 14912217, and National Natural Science Foundation of China (NSFC) grant with No.\ 13004007. 
S.Z.\ acknowledges support from the NAOJ ALMA Scientific Research Grant Code 2025-29B.
Q.G.\ acknowledges the support of the National Natural Science Foundation of China (NSFC) through grants No.\ 12403032 and the Shanghai Rising-Star Program (23YF1455600).
P.S.\ was partially supported by a Grant-in-Aid for Scientific Research (KAKENHI Number JP23H01221) of JSPS.
\end{acknowledgments}

\facilities{ALMA, SMA}
\software{CASA \citep{casa2007}, Astropy \citep{Astropy2013}, Emcee \citep{Foreman-Mackey2013}}

\appendix

\section{Integrated intensity maps of the 40 molecular lines used in NMF}\label{app:moment0}

\begin{figure*}[!thpb]
\centering
\includegraphics[width=0.9\textwidth]{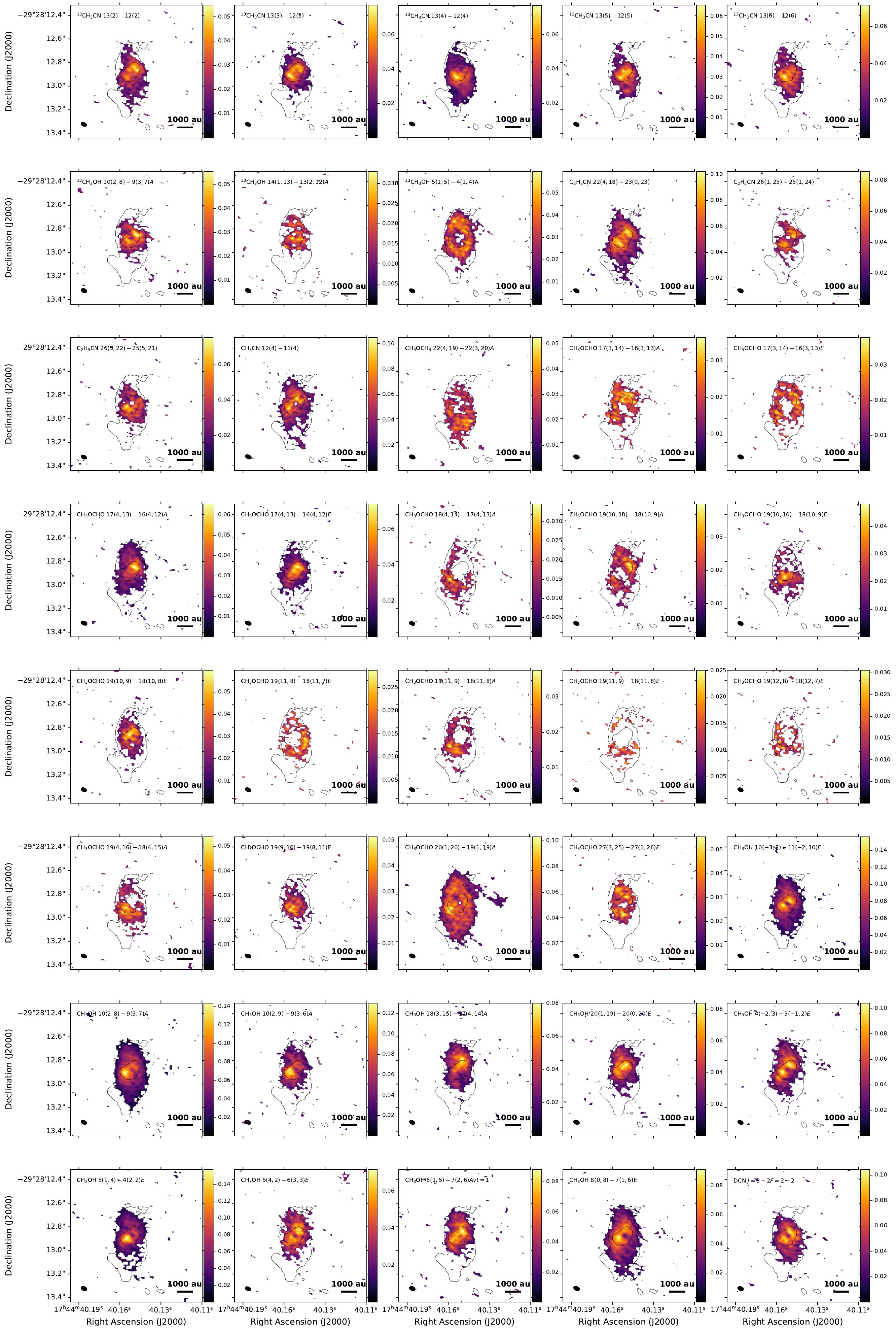}
\caption{Integrated intensity maps of the 40 molecular lines.
}
\label{fig:moment0_NMF}
\end{figure*}

\autoref{fig:moment0_NMF} presents the integrated intensity maps of the lines used in NMF analysis with a threshold of 3$\sigma$, including CH$_3$OCHO, CH$_3$OH, $^{13}$CH$_3$OH, CH$_3$CN, $^{13}$CH$_3$CN, and C$_2$H$_5$CN. The black contours indicate the continuum emission at levels of [10, 50]$\times\sigma$, where the RMS noise is $\sigma = 18~\mu\mathrm{Jy\,beam^{-1}}$. 

\section{NMF contribution of Sgr c disk}\label{app:nmf_fractions}

\begin{figure*}[!thpb]
\centering
\includegraphics[width=0.8\textwidth]{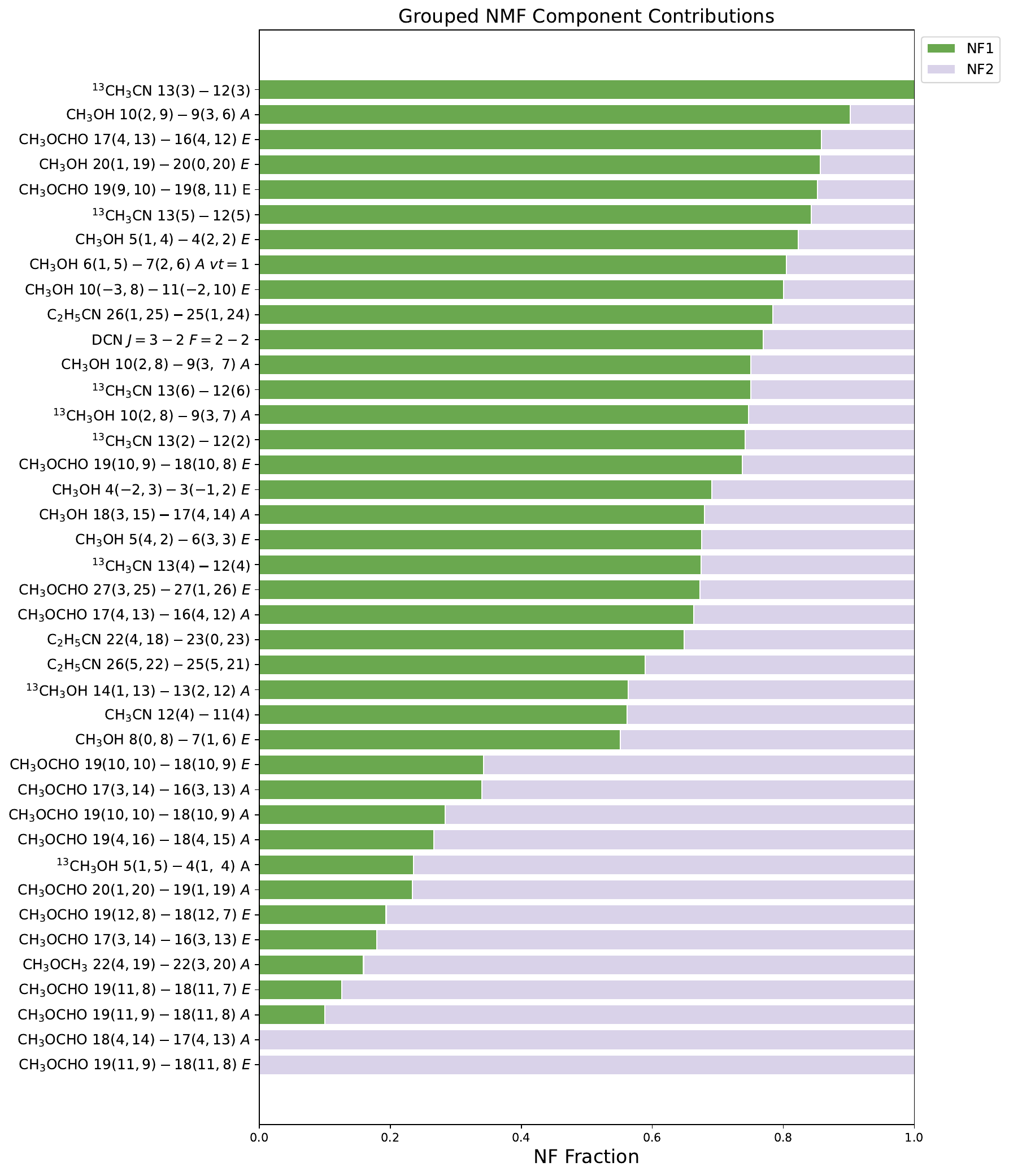
}
\caption{NMF contribution of of the molecular lines in the Sgr~C disk.
}
\label{fig:NMF_contribution}
\end{figure*}

 \autoref{fig:NMF_contribution} summarizes the morphological similarity of a specific spectral line emission with one of the two components in the NMF analysis. The quantitative NMF contribution of each line provides a clear chemical classification of these lines. 

\section{corner map}\label{app:corner_map}
\begin{figure*}[!thpb]
\centering
\includegraphics[width=0.8\textwidth]{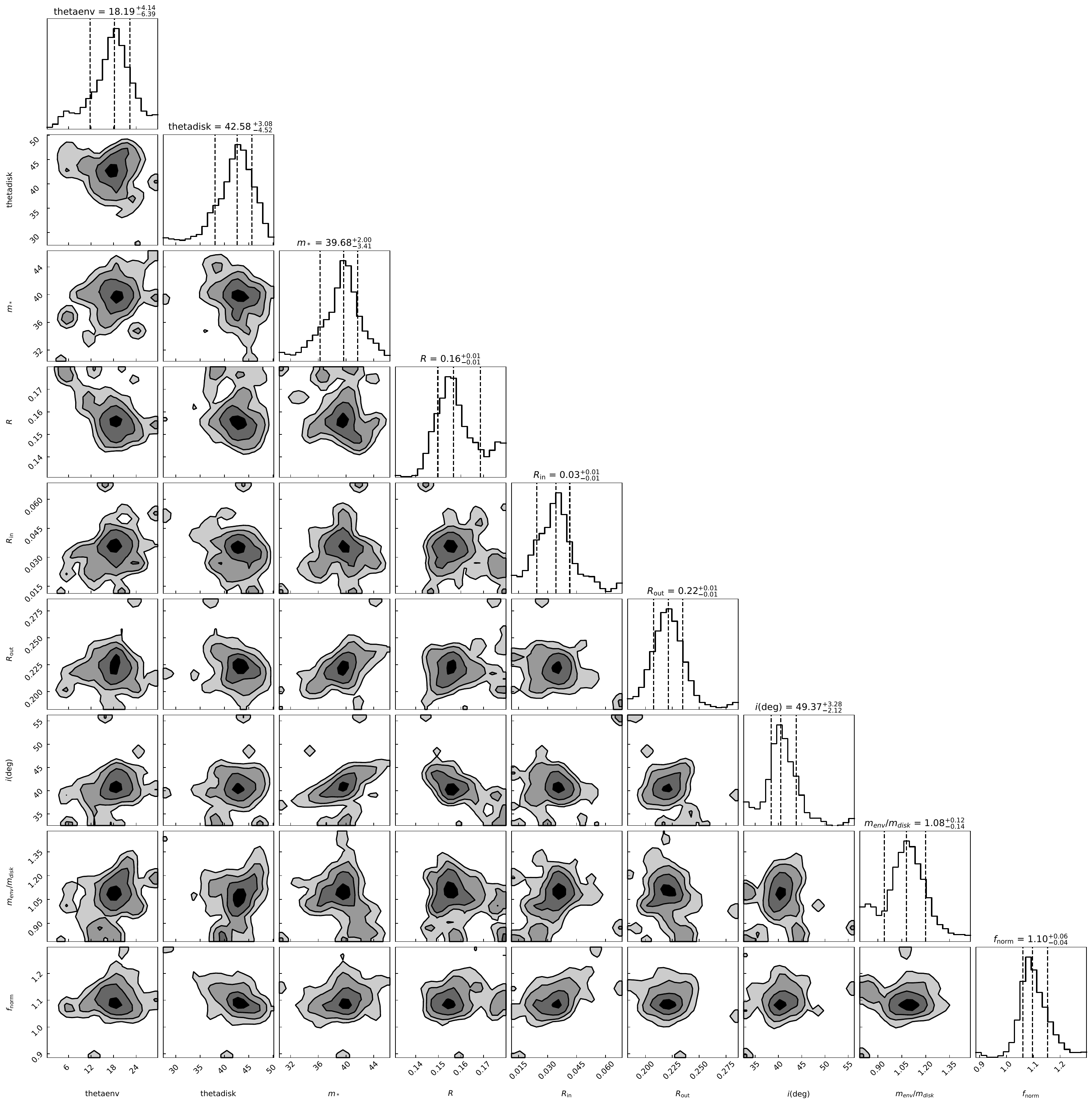}
\caption{Corner map of the \textit{emcee} fitting.
}
\label{fig:corner map}
\end{figure*}

The best-fit results from \textit{emcee} are presented in the corner map in \autoref{fig:corner map}.

\section{PV diagram across the Sgr~C disk}\label{app: PV}

\begin{figure*}[!thpb]
\centering
\includegraphics[width=0.45\textwidth]{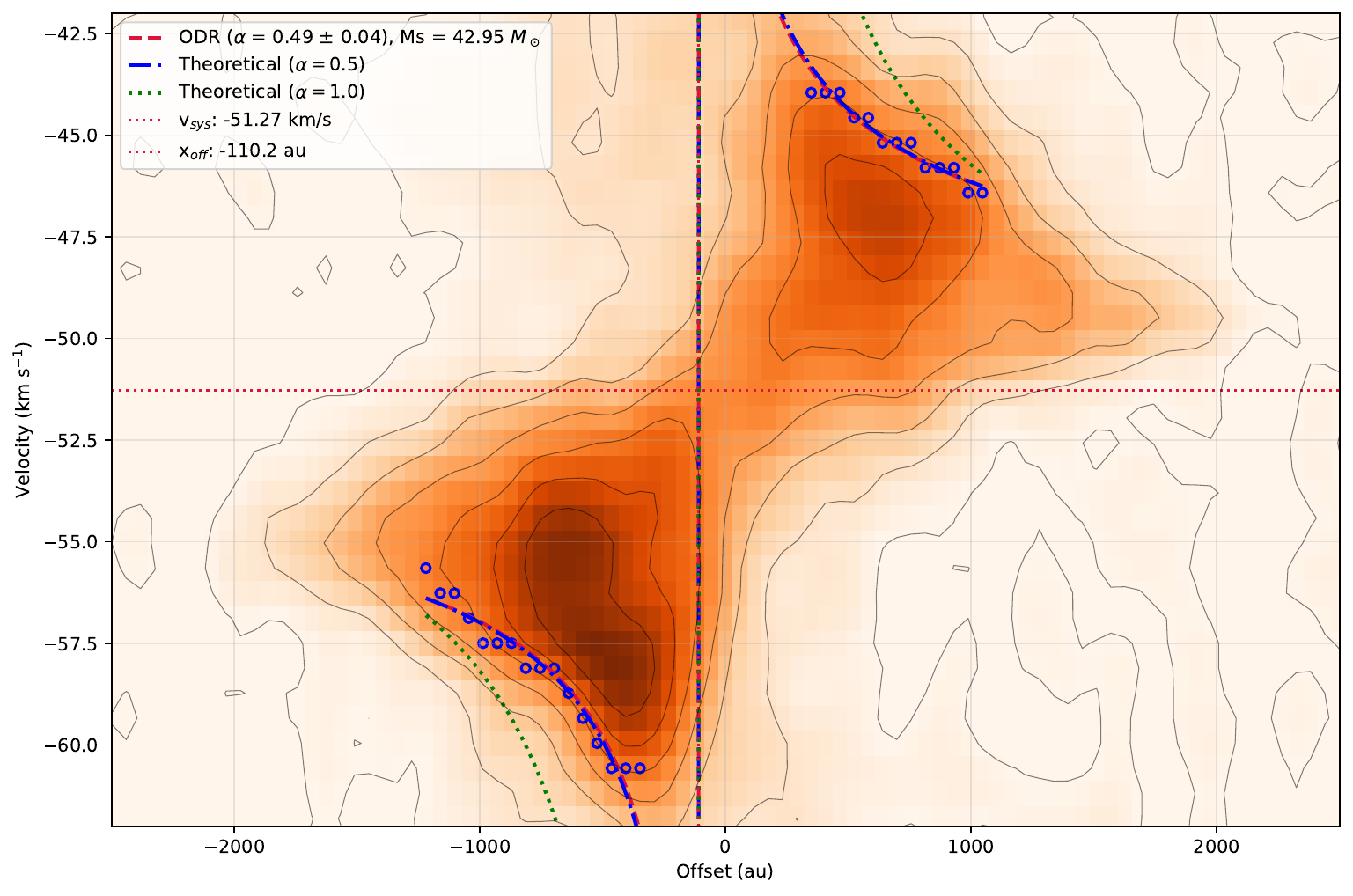}
\includegraphics[width=0.45\textwidth]{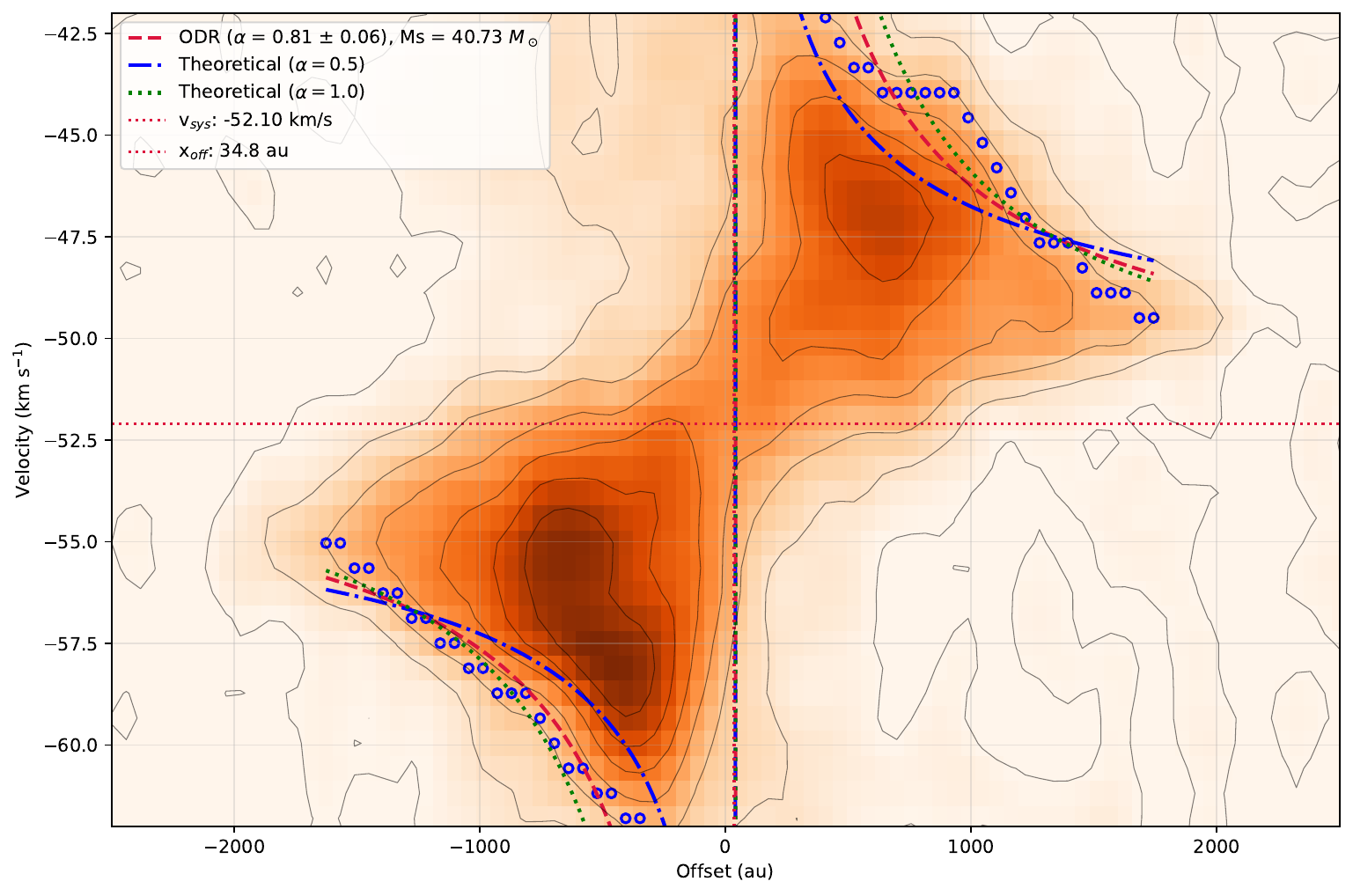
}
\caption{Position-velocity diagram of the stacked $\mathrm{CH_3OCHO}$ emission extracted along the visually determined major axis ($42^{\circ}$ north-to-west) of the Sgr~C disk. The data are fitted using Orthogonal Distance Regression at two different signal-to-noise ratio thresholds. \textbf{Left:} The $16\sigma$ threshold isolates the inner high-velocity emission ($< 1300$~au), yielding a power-law index of $\alpha = 0.49 \pm 0.04$. \textbf{right:} The $9\sigma$ threshold incorporates extended emission up to 1800~au, revealing a steeper index of $\alpha = 0.81 \pm 0.06$. Both fits suggest a central dynamical mass of $\sim 40~M_{\odot}$.}
\label{fig:PV_disk}
\end{figure*}

We analyze the position-velocity diagram extracted along the major axis (42$^{\circ}$ north-to-west) of the Sgr~C disk to constrain its dynamical properties and central mass. The major axis is derived in \autoref{subsec: model lines}. The line-of-sight velocity as a function of the spatial offset $x$ is described by the relation: 
\begin{equation}
    v(x) = v_{\mathrm{sys}} + \text{sign}(x - \theta_{\mathrm{off}}) a (|x - \theta_{\mathrm{off}}| / R_0)^{-\alpha},
    \label{eq:PV}
\end{equation}
where $a$ represents the projected rotation velocity at a fixed reference radius $R_0$, $v_{\mathrm{sys}}$ is the systemic velocity, $\theta_{\mathrm{off}}$ is the spatial offset of the central source, and $\alpha$ is the power-law index determining the velocity gradient. We determine the optimal parameters using Orthogonal Distance Regression (ODR) to account for uncertainties in both the positional offsets and the measured velocities, weighting the data points by the inverse square of their respective errors. Once the best-fit projected velocity amplitude $a$ is obtained, we derive the intrinsic deprojected rotation velocity $v_{\mathrm{rot}} = a / \sin(i)$, adopting a disk inclination angle of $i=50^{\circ}$. Finally, under the assumption that the gas kinematics are gravitationally bound and dominated by the potential of the central source, the central dynamical mass is calculated via $M_* = R_0 v_{\mathrm{rot}}^2 / G$, where $G$ is the gravitational constant. We set $R_0 = 1300$~au, which is the radius of the inner Keplerian disk derived in \autoref{subsec: model lines}.

To evaluate the dependence of the dynamics on the radial scale, we perform ODR fitting on velocity structures extracted using two different signal-to-noise ratio clipping thresholds (see \autoref{fig:PV_disk}). As shown in the PV diagram using a $16\sigma$ extraction threshold, the data points predominantly trace the inner, high-velocity emission within a radius of $1300~\mathrm{au}$. The ODR fit to this inner region yields a power-law index of $\alpha = 0.49 \pm 0.04$. This result matches the theoretical expectation for a rotationally supported purely Keplerian accretion disk ($\alpha = 0.5$). Based on this Keplerian profile and assuming an inclination angle of $50^{\circ}$, we derive a central dynamical mass of $M_* \approx 43.0~M_{\odot}$. In contrast, when a lower threshold $9\sigma$ is adopted, the extracted data points trace the more extended, lower-velocity emission to $1800~\mathrm{au}$. The best-fit power-law index for this extended structure steepens to $\alpha = 0.81 \pm 0.06$, representing an intermediate state between Keplerian rotation ($\alpha = 0.5$) and rotation with conserved specific angular momentum ($\alpha = 1.0$). \autoref{tab:odr_fit_results} presents the best-fit result of parameters.

This steepening in the outer radii suggests a transition from the inner Keplerian disk to an infalling envelope. In the outer envelope, the gas tends to conserve its angular momentum as it spirals inward before settling onto the rotationally supported disk. Despite this steepened gradient, the estimated central mass, $M_* \approx 40.7~M_{\odot}$, remains to be consistent with the result from the PPV dynamical modeling in \autoref{subsec: model lines}. Together, these two limiting cases robustly confirm that the Sgr~C disk system hosts a central high-mass protostar with $M_* \sim 40~M_{\odot}$.

\begin{table}[htbp]
\centering
\caption{ODR Fitting Results of the PV Diagram at Different S/N Thresholds}
\label{tab:odr_fit_results}
\begin{tabular}{lccccc}
\hline\hline
Threshold & $\alpha$ & $M_*$ ($M_{\odot}$) & $v_{\mathrm{sys}}$ ($\mathrm{km\,s^{-1}}$) & $x_{\mathrm{off}}$ ($\mathrm{au}$) & Dynamical State \\
\hline
$16\sigma$ & $0.49 \pm 0.04$ & 42.95 & $-$51.27 & $-$110.2 & Keplerian Disk \\
$9\sigma$  & $0.81 \pm 0.06$ & 40.73 & $-$52.10 & 34.8   & Infalling Envelope Transition \\
\hline
\end{tabular}
\tablecomments{The dynamical mass $M_*$ is estimated assuming an inclination angle of $50^{\circ}$.}
\end{table}

\bibliography{sgrcdisk}{}
\bibliographystyle{aasjournalv7}

\end{CJK}
\end{document}

%% file: authors.tex
\author[0009-0008-8439-8488]{Jixiang Weng(翁吉祥)}
\email[show]{wengjixiang@shao.ac.cn}
\affiliation{Shanghai Astronomical Observatory, Chinese Academy of Sciences, 80 Nandan Road, Shanghai 200030, People's Republic of China}
\affiliation{School of Astronomy and Space Sciences, University of Chinese Academy of Sciences,\\
No.\ 19A Yuquan Road, Beijing 100049, People's Republic of China}

\author[0000-0003-2619-9305]{Xing Lu(吕行)}
\email[show]{xinglu@shao.ac.cn}
\affiliation{State Key Laboratory of Radio Astronomy and Technology, Shanghai Astronomical Observatory, Chinese Academy of Sciences, \\
80 Nandan Road, Shanghai 200030, People's Republic of China}

\author[0000-0002-8691-4588]{Yu Cheng(程宇)}
\email{ycheng.astro@gmail.com}
\affiliation{National Astronomical Observatory of Japan, 2-21-1 Osawa, Mitaka, Tokyo, 181-8588, Japan}

\author[0000-0001-6858-1006]{Hongping Deng(邓洪平)}
\email{hpdeng353@shao.ac.cn}
\affiliation{Shanghai Astronomical Observatory, Chinese Academy of Sciences, 80 Nandan Road, Shanghai 200030, People's Republic of China}

\author[0000-0001-7573-0145]{Xiaofeng Mai(麦晓枫)}
\email{maixf@shao.ac.cn}
\affiliation{Shanghai Astronomical Observatory, Chinese Academy of Sciences, 80 Nandan Road, Shanghai 200030, People's Republic of China}

\author[0000-0002-8389-6695]{Suinan Zhang(张遂楠)}
\email{suinan.zhang@gmail.com}
\affiliation{Department of Earth and Planetary Sciences, Institute of Science Tokyo, Meguro, Tokyo, 152-8551, Japan}
\affiliation{National Astronomical Observatory of Japan, 2-21-1 Osawa, Mitaka, Tokyo, 181-8588, Japan}

\author[0000-0002-6073-9320]{Cara Battersby}
\email{cbattersby@gmail.com}
\affiliation{Department of Physics, University of Connecticut, 196A Auditorium Road, Unit 3046, Storrs, CT 06269, USA}

\author[0000-0001-6431-9633]{Adam Ginsburg}
\email{adamginsburg@ufl.edu}
\affiliation{Department of Astronomy, University of Florida, PO Box 112055, Gainesville, FL 32611, USA}

\author[0000-0001-8782-1992]{Elisabeth A.C. Mills}
\email{eacmills@ku.edu}
\affiliation{Department of Physics and Astronomy, University of Kansas, 1251 Wescoe Hall Drive, Lawrence, KS 66045, USA}

\author[0000-0001-7511-0034]{Yichen Zhang}
\email{yczhang.astro@gmail.com}
\affiliation{Department of Astronomy, School of Physics and Astronomy, Shanghai Jiao Tong University, 800 Dongchuan Road, Shanghai 200240, China}
\affiliation{State Key Laboratory of Dark Matter Physics, School of Physics and Astronomy, Shanghai Jiao Tong University, Shanghai 200240, China}
\affiliation{Key Laboratory for Particle Astrophysics and Cosmology (MOE) / Shanghai Key Laboratory for Particle Physics and Cosmology, Shanghai 200240, China}

\author[0000-0002-5286-2564]{Tie Liu}
\email{liutiepku@gmail.com}
\affiliation{State Key Laboratory of Radio Astronomy and Technology, Shanghai Astronomical Observatory, Chinese Academy of Sciences, \\
80 Nandan Road, Shanghai 200030, People's Republic of China}

\author[0000-0003-2133-4862]{Thushara Pillai}
\email{tpillai.astro@gmail.com}
\affiliation{Haystack Observatory, Massachusetts Institute of Technology, 99 Millstone Road, Westford, MA 01886, USA}

\author[0000-0002-4154-4309]{Xindi Tang}
\email{tangxindi@xao.ac.cn}
\affiliation{Xinjiang Astronomical Observatory, Chinese Academy of Sciences, 830011 Urumqi, People's Republic of China}

\author[0000-0002-8250-6827]{Fernando Olguin}
\email{f.olguin.ch@gmail.com}
\affiliation{Center for Gravitational Physics, Yukawa Institute for Theoretical Physics, Kyoto University, Sakyo-ku, Kyoto 606-8502, Japan}
\affiliation{National Astronomical Observatory of Japan, 2-21-1 Osawa, Mitaka, Tokyo, 181-8588, Japan}

\author[0000-0003-2300-2626]{Hauyu Baobab Liu}
\email{baobabyoo@gmail.com}
\affiliation{Department of Physics, National Sun Yat-Sen University, No. 70, Lien-Hai Road, Kaohsiung City 80424, Taiwan, ROC}

\author[0000-0003-2384-6589]{Qizhou Zhang}
\email{qzhang@cfa.harvard.edu}
\affiliation{Center for Astrophysics $\vert{}$ Harvard $\&$ Smithsonian, 60 Garden Street, Cambridge, MA 02138, USA}

\author[0000-0003-1275-5251]{Shanghuo Li}
\email{shli@nju.edu.cn}
\affiliation{School of Astronomy and Space Science, Nanjing University, 163 Xianlin Avenue, Nanjing 210023, People's Republic of China}
\affiliation{Key Laboratory of Modern Astronomy and Astrophysics (Nanjing University), Ministry of Education, Nanjing 210023, People's Republic of China}

\author[0000-0003-3144-1952]{Guang-Xing Li}
\email{ligx.ngc7293@gmail.com}
\affiliation{South-Western Institute for Astronomy Research, Yunnan University, Kunming, People’s Republic of China}

\author[0000-0002-7125-7685]{Patricio Sanhueza}
\email{patosanhueza@gmail.com}
\affiliation{Department of Astronomy, School of Science, The University of Tokyo, 7-3-1 Hongo, Bunkyo, Tokyo 113-0033, Japan}

\author[0009-0005-9867-6723]{Yunfan Jiao}
\email{Yunfan.astro@gmail.com}
\affiliation{Shanghai Astronomical Observatory, Chinese Academy of Sciences, 80 Nandan Road, Shanghai 200030, People's Republic of China}
\affiliation{School of Astronomy and Space Sciences, University of Chinese Academy of Sciences,\\
No.\ 19A Yuquan Road, Beijing 100049, People's Republic of China}

\author[0000-0002-2826-1902]{Qilao Gu}
\email{qlgu@shao.ac.cn}
\affiliation{Shanghai Astronomical Observatory, Chinese Academy of Sciences, 80 Nandan Road, Shanghai 200030, People's Republic of China}

\author[0009-0002-1465-1958]{Kai Smith}
\email{smith.kai682@ku.edu}
\affiliation{Department of Physics and Astronomy, University of Kansas, 1251 Wescoe Hall Drive, Lawrence, KS 66045, USA}

\author[0000-0001-7817-1975]{YanKun Zhang}
\email{zhangyankun@shao.ac.cn}
\affiliation{Shanghai Astronomical Observatory, Chinese Academy of Sciences, 80 Nandan Road, Shanghai 200030, People's Republic of China}

\author[0009-0000-5899-4376]{Tianning Lyu}
\email{lvtianning@shao.ac.cn}
\affiliation{Shanghai Astronomical Observatory, Chinese Academy of Sciences, 80 Nandan Road, Shanghai 200030, People's Republic of China}
\affiliation{School of Astronomy and Space Sciences, University of Chinese Academy of Sciences,\\
No.\ 19A Yuquan Road, Beijing 100049, People's Republic of China}

\author[0000-0003-3540-8746]{Zhiqiang Shen}
\email{zshen@shao.ac.cn}
\affiliation{Shanghai Astronomical Observatory, Chinese Academy of Sciences, 80 Nandan Road, Shanghai 200030, People's Republic of China}